\documentclass[prd,preprint,floatfix,nofootinbib,preprintnumbers,superscriptaddress,showkeys]{revtex4} 
\pdfoutput=1
\usepackage[caption=false]{subfig}
\usepackage{amsmath}
\usepackage{amsfonts}
\usepackage{amssymb}
\usepackage{float}
\usepackage{color}
\usepackage{graphicx}
\usepackage{graphics}
\usepackage{hyperref}
\usepackage{adjustbox,array}
\usepackage{enumitem} 
\hypersetup{}
\usepackage[utf8x]{inputenc} 
\usepackage{multirow}
\usepackage{booktabs}
\usepackage{mathrsfs}
\usepackage{diagbox}
\usepackage[english]{babel}
\usepackage[autostyle]{csquotes}

\definecolor{rosso}{cmyk}{0,1,1,0.4}
\definecolor{rossos}{cmyk}{0,1,1,0.55}
\definecolor{rossoc}{cmyk}{0,1,1,0.2}
\definecolor{blu}{cmyk}{1,1,0,0.3}
\definecolor{blus}{cmyk}{1,1,0,0.6}
\definecolor{bluc}{cmyk}{1,1,0,0.1}
\definecolor{verde}{cmyk}{0.92,0,0.59,0.25}
\definecolor{verdec}{cmyk}{0.92,0,0.59,0.15}
\definecolor{verdes}{cmyk}{0.92,0,0.59,0.4}

\AtBeginDocument{
\heavyrulewidth=.08em
\lightrulewidth=.05em
\cmidrulewidth=.03em
\belowrulesep=.65ex
\belowbottomsep=0pt
\aboverulesep=.4ex
\abovetopsep=0pt
\cmidrulesep=\doublerulesep
\cmidrulekern=.5em
\defaultaddspace=.5em
}

\hypersetup{
 colorlinks=true,
 linkcolor=verdec,
 urlcolor=verdec,
 citecolor=verdec
}

\newcommand{ \eq}[1]{Eq.~(\ref{#1})}

\newcommand{\gsim}{\gtrsim}
\newcommand{\lsim}{\lesssim}

\newcommand{\lf}{\left(}
\newcommand{\ri}{\right)}

\newcommand{\nn}{\nonumber}
\newcommand{\rd}{\partial}
\newcommand{\drdmu}{\!\stackrel{\!\!\leftrightarrow}{\partial^\mu}\!}

\newcommand{\sqt}{\sqrt{2}}

\renewcommand{\lg}{\mathscr{L}} 

\newcommand{\mco}{\mathcal{O}}

\newcommand{\br}{\mathcal{B}}
\newcommand{\hc}{{\rm H.c.}}

\newcommand{\tot}{{\rm tot}}

\newcommand{\pb}{{\;{\rm pb}}}
\newcommand{\fb}{{\;{\rm fb}}}

\newcommand{\mev}{{\;{\rm MeV}}}
\newcommand{\gev}{{\;{\rm GeV}}}

\newcommand{\beq}{\begin{equation}}
\newcommand{\eeq}{\end{equation}}
\newcommand{\bea}{\begin{eqnarray}}
\newcommand{\eea}{\end{eqnarray}}
\newcommand{\barr}{\begin{array}}
\newcommand{\earr}{\end{array}}
\newcommand{\bc}{\begin{center}}
\newcommand{\ec}{\end{center}}
\newcommand{\bit}{\begin{itemize}}
\newcommand{\eit}{\end{itemize}}
\newcommand{\ben}{\begin{enumerate}}
\newcommand{\een}{\end{enumerate}}

\newcommand{\al}{\alpha}
\newcommand{\bt}{\beta}

\newcommand{\Dt}{\Delta}

\newcommand{\sg}{\sigma}
\newcommand{\es}{\epsilon}
\newcommand{\kp}{\kappa}

\newcommand{\gm}{\gamma}
\newcommand{\Gm}{\Gamma}
\newcommand{\lm}{\lambda}

\newcommand{\tauh}{\tau_{\rm h}}

\newcommand{\hsm}{{h_{\rm SM}}}
\newcommand{\ch}{H^\pm}

\newcommand{\wpm}{W^\pm}
\newcommand{\wmp}{W^\mp}
\newcommand{\mh}{m_{h}}
\newcommand{\mch}{M_{H^\pm}}
\newcommand{\mhh}{M_{H}}
\newcommand{\ma}{M_{A}}

\newcommand{\tb}{t_\beta}
\newcommand{\tbi}{\frac{1}{t_\beta}}
\newcommand{\cb}{c_\beta}
\renewcommand{\sb}{s_\beta}

\newcommand{\cba}{c_{\beta-\alpha}}
\newcommand{\sba}{s_{\beta-\alpha}}

\newcommand{\ttau}      {{\tau^+\tau^-}} 

\newcommand{\ttop}      {{t\bar{t}}}

\newcommand{\bb}      {{b \bar{b}}}
\newcommand{\ww}      {{W^+ W^-}}

\newcommand{\qq}      {{q \bar{q}}}

\newcommand{\met}      {{E_T^{\rm miss}}}

\newcommand{\ptll}      {p_T^{\ell^{(\rm lead)}}}

\definecolor{mint}{rgb}{0.24, 0.71, 0.54}

\newcommand{\textoverline}[1]{$\overline{\mbox{#1}}$}

\begin{document}

\title{\color{verdes} Comprehensive study of the light charged Higgs boson \\
in the type-I two-Higgs-doublet model}
\author{ Kingman Cheung}
\email{cheung@phys.nthu.edu.tw}
\address{Department of Physics, Konkuk University, Seoul 05029, Republic of Korea}
\address{Department of Physics, National Tsing Hwa University, Hsinchu 300, Taiwan}
\address{Center for Theory and Computation, National Tsing Hua University,
  Hsinchu 300, Taiwan}
\author{Adil Jueid}
\email{adiljueid@kias.re.kr}
\address{Quantum Universe Center, Korea Institute for Advanced Study, Seoul 02455, Republic of Korea}
\address{Department of Physics, Konkuk University, Seoul 05029, Republic of Korea}
\author{Jinheung Kim}
\email{jinheung.kim1216@gmail.com}
\address{Department of Physics, Konkuk University, Seoul 05029, Republic of Korea}
\author{Soojin Lee}
\email{soojinlee957@gmail.com}
\address{Department of Physics, Konkuk University, Seoul 05029, Republic of Korea}
\author{Chih-Ting Lu}
\email{timluyu@gmail.com}
\address{School of Physics, Korea Institute for Advanced Study, Seoul 02455, Republic of Korea}
\address{Department of Physics and Institute of Theoretical Physics, Nanjing Normal University, Nanjing, 210023, China}
\author{Jeonghyeon Song}
\email{jhsong@konkuk.ac.kr}
\address{Department of Physics, Konkuk University, Seoul 05029, Republic of Korea}

\begin{abstract}
In the type-I two-Higgs-doublet model, existing theoretical and experimental constraints still permit the light charged Higgs boson with a mass below the top quark mass. We present a complete roadmap for the light charged Higgs boson at the LHC through the comprehensive phenomenology study, focusing on the normal scenario where the lighter \textit{CP}-even Higgs boson is the observed Higgs boson. In type-I, it is challenging to simultaneously accommodate the light mass of the charged Higgs boson and the constraints from theory, electroweak precision data, Higgs data, $b\to s  \gamma$, and direct search bounds. Consequently, the parameter space is extremely curtailed, which predicts somewhat definite phenomenological signatures. We find that the mass of the pseudoscalar Higgs boson, $M_A$, is the most crucial factor in the phenomenology of the charged Higgs boson. If $M_A$ is light, the charged Higgs boson decays mainly into $A W^\pm$. When $M_A$ is above the $A W^\pm$ threshold, the dominant decay mode is into $ \tau^\pm \nu$. Over the whole viable parameter space, we study all the possible production and decay modes of charged Higgs bosons at the LHC, and suggest three efficient channels: (i) $pp \to H^+ H^-\to [\tau \nu] [\tau \nu]$; (ii) $pp \to HA/HH/AA \to H^\pm W^\mp H^\pm W^\mp \to [\tau \nu] [\tau \nu] W W$; (iii) $pp \to H^+ H^-\to [b \bar b W ]  [b \bar b W ] $. Based on the sophisticated signal-background analyses including detector simulation, we showed that the significance of the first final state is large, that of the second one is marginal around three, but the third one suffers from huge $t\bar t$ related backgrounds. 
\end{abstract}

\vspace{1cm}

\preprint{KIAS-Q22002}
\maketitle
\tableofcontents

\section{Introduction}

The discovery of the Higgs boson at the LHC in 2012~\cite{ATLAS:2012yve,CMS:2012qbp}
is a triumph achieved through cooperation between the theoretical and experimental communities in particle physics.
Despite the completion of the standard model (SM),
however, we still long for the next milestone to progress toward the final theory of the Universe,
as facing the baffling questions
such as the naturalness problem, the fermion mass hierarchy, the origin of \textit{CP} violation in the quark sector,
the baryogenesis, the non-zero neutrino masses, and the identity of dark matter.
Since 2012, the ATLAS and CMS collaborations have searched hard for the same success as the observed Higgs boson,
a dramatic resonance bump in invariant mass distribution, but not achieved any success so far.
A new direction of research arises in the framework of the SM effective
field theory~\cite{Buchmuller:1985jz}
where we systematically characterize the experimental deviations from the SM
predictions without specifying the UV physics.

Nevertheless, direct searches for new particles should continue
because they can explicitly reveal an essential aspect of the new physics (NP) theory.
Many NP models have an extended Higgs sector.
When additional Higgs doublets, triplets, or higher representations are included,
a distinguished new particle is the charged Higgs boson $\ch$.
If $\ch$ is light at a mass below the top quark mass, the implication on the UV theory shall be further profound.
From this perspective, 
we consider the light charged Higgs boson in the two-Higgs-doublet model (2HDM)~\cite{Aoki:2009ha,Branco:2011iw,Craig:2013hca},
which accommodates five Higgs bosons,
\textit{CP}-even
neutral $h$ and $H$ ($\mh<\mhh$),
\textit{CP}-odd neutral $A$, and a pair of charged $\ch$.
The charged Higgs boson in type-II and type-Y is tightly constrained to be as heavy as $\mch\gsim 800\gev$
due to the measurements of the inclusive weak radiative $B$-meson decay into
$s \gm$~\cite{Misiak:2020vlo}.
Only type-I and type-X can accommodate a light $\ch$.
We concentrate on type-I in this paper.
In type-I,  all the Yukawa couplings of $\ch$
are inversely proportional to $\tan\bt$, 
the ratio of two vacuum expectation values of two Higgs-doublet fields.
The decay branching ratios of $\ch$ into a fermion pair are proportional to the fermion mass,
which suggested the main search mode at the LHC as the production via
the decay of the top quark $t \to b H^+$, followed by the decay $\ch\to \tau\nu$.
Both the ATLAS and CMS collaborations have analyzed this
mode~\cite{ATLAS:2018gfm,Sirunyan:2019hkq},
presenting the upper bound  on 
$\br(t\to b \ch)\times \br(\ch\to \tau\nu)$. 
The absence of new signal demands large $\tan\beta$, e.g.,
$\tan\bt \gsim 10$ for $\mch=110\gev$,
which highly suppresses $\br(t\to b \ch)$ below $\mco(10^{-4})$.
Thereupon we come to question
whether $t\to b \ch$ is indeed the golden mode for the light charged Higgs boson in type-I.

\begin{table}
\begin{tabular}{|l||c|c|c|c|c|}
\hline
\diagbox{Production}{Decay} & $[\tau^\pm \nu]$ & $[c b]$ & $[c s]$ & $[W^\pm \varphi^0/ A]$ & $[W^\pm \hsm/A]$\\
\hline\hline
\multirow{5}{*}{$t \to \ch b$} & type-I~\cite{Abbaspour:2018ysj,Sanyal:2019xcp} & 3HDM~\cite{Akeroyd:2018axd} &3HDM~\cite{Akeroyd:2018axd}  & IS type-I~\cite{Arhrib:2020tqk} & type-I~\cite{Arhrib:2016wpw}\\
 & type-X~\cite{Demir:2018iqo,Sanyal:2019xcp} & & & N2HDM~\cite{Dermisek:2012cn} & \\ \cline{2-6}
 & ATLAS~\cite{ATLAS:2018gfm}& ATLAS~\cite{ATLAS:2021zyv} & ATLAS~\cite{ATLAS:2013uxj}& CMS~\cite{CMS:2019idx}&\\
 & CMS~\cite{Sirunyan:2019hkq}& CMS~\cite{CMS:2018dzl} & CMS~\cite{CMS:2015yvc,CMS:2020osd} &  &\\  \hline
\multirow{2}{*}{$W^{\pm *} \to \ch\varphi^0$} & \multirow{2}{*}{}& \multirow{2}{*}{} & \multirow{2}{*}{}& IS type-I~\cite{Arhrib:2017wmo} &\multirow{2}{*}{}\\ 
 & & & & Fermiphobic type-I~\cite{Mondal:2021bxa}  &\\ \hline
$W^{\pm *} \to \ch A$ & type-X~\cite{Kanemura:2011kx} &  & & IS type-I,X~\cite{Arhrib:2021xmc,Arhrib:2021yqf} &\\
\hline
$pp\to H^+ H^-$ &  &  & & IS type-I,X~\cite{Arhrib:2021xmc,Arhrib:2021yqf} &\\
\hline
$q b \to q' b \ch$ & MSSM~\cite{Moretti:1996ra} & & & & \\ \hline
$cs/cb \to \ch$ & type-III~\cite{Hernandez-Sanchez:2012vxa,Hernandez-Sanchez:2020vax}
& & & & \\ \hline 
$W^{\pm *}W^{\pm *} \to H^\pm H^\pm$ & $\br_{\tau\nu}=1$~\cite{Aiko:2019mww} & & & type-I,X~\cite{Arhrib:2019ywg} &\\
\hline
\end{tabular}
\caption{\label{table:literature}
Theoretical and experimental studies on a light charged Higgs boson in the 2HDM and 3HDM at the LHC,
classified according to the production and decay channels.
$\varphi^0$ denotes a \textit{CP}-even scalar boson with a mass below $125\gev$.
The theoretical model is also presented:
type-I, type-X, and type-III denote the type of 2HDM, ``IS" denotes the inverted scenario for $\hsm=H$ in the 2HDM,
and 3HDM is the three-Higgs-doublet model.
}
\end{table}

In this regard, other production and decay channels of the light charged Higgs boson in the 2HDM and 3HDM
at the LHC\footnote{Future colliders have been shown efficient 
  for production of a light $\ch$, such as future electron-proton colliders
  for $\ch$ in type-III 2HDM~\cite{Flores-Sanchez:2018dsr} and 
  three-Higgs-doublet
  model~\cite{Akeroyd:2018axd,Akeroyd:2019mvt}.} have been
studied recently.
We summarize the literature survey on the light charged Higgs boson\footnote{Some
unconventional decay channels of the heavy $\ch$ in the 2HDM have also been studied,
such as $\ch\to \wpm A$~\cite{Bahl:2021str}, $\ch\to \wpm\gm$~\cite{Song:2019aav},
and $H^\pm \to t\bar{b}$~\cite{Moretti:1999bw,Moretti:2016jkp,Arhrib:2018bxc}.} in Table \ref{table:literature},
specifying the theoretical model\footnote{In Ref.~\cite{Aiko:2019mww},
$\br_{\tau\nu} \equiv \br(\ch\to \tau^\pm \nu)=1$ is assumed without specifying the type of the 2HDM.}, the production channel, and the decay mode.
Here ``IS" stands for the inverted scenario where the observed Higgs boson at a mass of 125 GeV
is the heavier \textit{CP}-even $H$, while the light \textit{CP}-even Higgs boson, denoted
by $\varphi^0$ in Table \ref{table:literature},
has not been observed yet~\cite{Bernon:2015wef,Chang:2015goa,Arhrib:2017wmo,Arhrib:2020tqk,Jueid:2021avn,Arhrib:2021xmc,Arhrib:2021yqf}.
The studies in Table \ref{table:literature} reveal some aspects of 
the characteristics of the light charged Higgs boson in type-I,
but not the whole, because they focus on one or two specific channels.
In addition, many studies are based on some conditions 
such as the Higgs alignment limit for the SM-like Higgs boson~\cite{Carena:2013ooa,Celis:2013rcs,Cheung:2013rva,Bernon:2015qea,Chang:2015goa,Das:2015mwa,Kanemura:2021dez}
and the mass degeneracy of new Higgs bosons for the electroweak precision
data~\cite{Kanemura:2011sj,Chang:2015goa,Chen:2019pkq}.
But imposing the conditions could have interfered with the observation at the LHC.
In order not to miss the light charged Higgs boson, therefore,
we need a full roadmap over the whole viable parameter space of type-I.
Then, it is essential to investigate all the possible production and decay modes 
as well as the optimal and representative channel for each region of the parameter space.

To achieve the goal,
we will explore the entire parameter space of type-I with the light $\ch$,
and obtain the phenomenologically viable parameters.
As shall be shown,
imposing light $\mch$ restricts the model severely.
In turn, 
the model parameters are strongly correlated with each other.
Making the most of this feature,
we pursue the efficient discovery channels of the light charged Higgs bosons at the LHC,
which have definite signal rates throughout the allowed parameter space, i.e.,
weak dependence on the model parameters.
As shall be shown, the \textit{pair production} of the light charged Higgs boson serves our purpose.
Based on these results,
we will suggest three channels to cover the whole parameter space effectively:
 (i) $pp \to H^+ H^-\to [\tau \nu] [\tau \nu]$; (ii) $pp \to HA/HH/AA \to H^\pm W^\mp H^\pm W^\mp \to [\tau \nu] [\tau \nu] W W$;
(iii) $pp \to H^+ H^-\to [b \bar b W ]  [b \bar b W ] $.
Using sophisticated signal-background analysis techniques with the detector simulation, the LHC discovery potentials of the proposed channels are to be rigorously obtained.
These are our new contributions.

The paper is organized as follows.
In Sec.~\ref{sec:review}, we briefly review the type-I 2HDM with \textit{CP} invariance and softly broken $Z_2$ parity.
In Sec.~\ref{sec:scan}, we present the results of random scans 
by placing the theoretical and experimental constraints for $\mch=110,140\gev$.
The characteristic features of the allowed parameters are to be discussed,
including the branching ratios of the new Higgs bosons.
Section \ref{sec:production} deals with the production channels of a light $\ch$ at the LHC.
After finding all the possible signals,
we suggest three main processes which can cover the allowed parameter space.
In Sec.~\ref{sec:results},
we perform the signal-to-background analysis for
$pp \to[\tau\nu][\tau\nu]$,
$pp \to [\tau\nu][\tau\nu]WW$, and
$pp \to [bbW][bbW]$  at the HL-LHC.
Conclusions are given in Sec.~\ref{sec:conclusions}.

\section{Review of type-I 2HDM}
\label{sec:review}

The 2HDM accommodates two complex $SU(2)_L$ Higgs doublet scalar fields, $\Phi_1$ and $\Phi_2$~\cite{Branco:2011iw}:
\bea
\label{eq:phi:fields}
\Phi_i = \left( \begin{array}{c} w_i^+ \\[3pt]
\dfrac{v_i +  h_i + i \eta_i }{ \sqrt{2}}
\end{array} \right), \quad i=1,2,
\eea
where $v_{1}$ and $v_2$ are the nonzero vacuum expectation values of
$\Phi_1$ and $\Phi_2$, 
respectively.
The ratio of $v_2$ to $v_1$ defines the mixing angle $\beta$ by
$\tan \beta =v_2/v_1$.
In what follows, we use the simplified notation of
  $s_x=\sin x$, $c_x = \cos x$, and $t_x = \tan x$.
The electroweak symmetry is broken by $v =\sqrt{v_1^2+v_2^2}=246\gev $.
The flavor-changing-neutral-current (FCNC)
at tree level is prevented by a discrete $Z_2$ symmetry,
under which $\Phi_1 \to \Phi_1$
and $\Phi_2 \to -\Phi_2$~\cite{Glashow:1976nt,Paschos:1976ay}.
Then the most general and renormalizable scalar potential with \textit{CP} invariance is
\bea
\label{eq:VH}
V = && m^2 _{11} \Phi^\dagger _1 \Phi_1 + m^2 _{22} \Phi^\dagger _2 \Phi_2
-m^2 _{12} ( \Phi^\dagger _1 \Phi_2 + \hc) \\ \nn
&& + \frac{1}{2}\lambda_1 (\Phi^\dagger _1 \Phi_1)^2
+ \frac{1}{2}\lambda_2 (\Phi^\dagger _2 \Phi_2 )^2
+ \lambda_3 (\Phi^\dagger _1 \Phi_1) (\Phi^\dagger _2 \Phi_2)
+ \lambda_4 (\Phi^\dagger_1 \Phi_2 ) (\Phi^\dagger _2 \Phi_1) \\ \nn
&& + \frac{1}{2} \lambda_5
\left[
(\Phi^\dagger _1 \Phi_2 )^2 +  \hc
\right],
\eea
where the $m^2 _{12}$ term softly breaks the $Z_2$ parity.
The model accommodates five physical Higgs bosons, the light \textit{CP}-even scalar $h$,
the heavy \textit{CP}-even scalar $H$, the \textit{CP}-odd pseudoscalar $A$,
and a pair of charged Higgs bosons $H^\pm$.
The relations of the physical Higgs bosons with the weak eigenstates
in Eq.~(\ref{eq:phi:fields}) via two mixing angles $\al$ and $\bt$ 
are referred to Ref.~\cite{Song:2019aav}.

The SM Higgs boson $\hsm$ is
\bea
\hsm = \sba h + \cba H.
\eea
We take the normal scenario where the observed Higgs boson is $h$.
Of special importance is the Higgs alignment limit where $h=\hsm$.
When $\sba=1$, $H \to WW/ZZ$, $A \to Z h$, and $\ch\to W^{\pm (*)}h$ are prohibited at tree level,
but the exotic Higgs decay $h \to AA$ is allowed if $A$ is light enough.
In this paper, we do not make any assumption on the model parameters.
Only the theoretical and experimental constraints determine the phenomenology.

We take the physical parameter basis of 
\bea
\label{eq:model:parameters}
\left\{\mh,\quad \mch, \quad\mhh,\quad \ma,\quad m_{12}^2,\quad \tb,\quad\sba \right\},
\eea
where $\beta-\al\in[0,\pi]$.\footnote{The public codes such as \textsc{2HDMC}~\cite{Eriksson:2009ws},
\textsc{HiggsSignals}~\cite{Bechtle:2020uwn}, and \textsc{HiggsBounds}~\cite{Bechtle:2020pkv}
take the range of $(\beta-\al) \in [-\pi/2,\pi/2]$,
but most of the theoretical studies adopt the convention of $\sba>0$.
For the immediate comparison with other theoretical studies, 
we present the results in the positive $\sba$ scheme:
if $\sba^{\textsc{2HDMC}}<0$, $(\beta-\al)=(\beta-\al)^{\textsc{2HDMC}} + \pi$. }
The quartic couplings  in the scalar potential play an essential role in satisfying the theoretical constraints.
In terms of the model parameters,
they are given as~\cite{Gunion:2002zf,Kanemura:2011sj}
\bea
\lm_1 &=& \frac{1}{v^2 }
\left[ m_h^2 \lf \sba - \cba \tb \ri ^2 
+  \mhh^2 \lf \sba\tb+\cba \ri^2
- M^2 \tb^2 \right],
\\[3pt] \nn
\lm_2 &=& \frac{1}{v^2 }
\left[ \mh^2 \lf\sba+ \frac{\cba}{\tb} \ri^2 - \frac{M^2 }{\tb^2}  +\mhh^2 \lf \frac{\sba}{\tb} - \cba\ri^2 
\right],
\\[3pt] \nn
\lm_3 &=&  \frac{1}{v^2}
\left[(\mh^2-\mhh^2) \left\{ \sba^2-\sba\cba\lf \tb-\frac{1}{\tb} \ri -\cba^2\right\} + 2 \mch^2 - M^2 
\right],
\\[3pt] \nn
\lm_4 &=& \frac{1}{v^2}
\left[
M^2+\ma^2-2 \mch^2
\right],
\\[3pt] \nn
\lm_5 &=& \frac{1}{v^2}
\left[
M^2-\ma^2
\right],
\eea
where $M^2=m_{12}^2/(\sb\cb)$.

The gauge couplings of the Higgs bosons are described by
\bea
\label{eq:gauge:Lg}
\lg_{\rm gauge} &=&
\Big( g m_W W^\dagger_\mu W^\mu
+
\dfrac{1}{2} g_Z m_Z Z_\mu Z^\mu
\Big)
\Big(
\sba h + \cba H
\Big)
\\[3pt] \nn && + \dfrac{g}{2} i
\left[
W_\mu^+ \lf \cba h -\sba H \ri \drdmu H^-   - H.c.
\right]
- \dfrac{g}{2}
\left[
W_\mu^+  H^- \drdmu A  + H.c.
\right]
\\[3pt] \nn &&
+ i
\left\{e A_\mu + \dfrac{g_Z}{2}(s_W^2-c_W^2) Z_\mu\right\} H^+ \drdmu H^-
+ \dfrac{g_Z}{2}  Z_\mu
\left[
\cba A \drdmu h -\sba A \drdmu H 
\right],
\eea
where $s_W =\sin\theta_W$, 
$g_Z =  g/c_W$, and $f \drdmu g \equiv  \left( f \,\rd^\mu g - g\, \rd^\mu f\right)$.
The Yukawa couplings to the SM fermions are defined by
\bea
\lg_{\rm Yuk} &=&
- \sum_f 
\lf 
\frac{m_f}{v} \kp_f \bar{f} f h + \frac{m_f}{v} \xi^H_f \bar{f} f H
-i \frac{m_f}{v} \xi^A_f \bar{f} \gm_5 f A
\ri
\\ \nn &&
- 
\left\{
\dfrac{\sqrt2V_{ud}}{v } H^+  \overline{u}
\left(m_u \xi^A_u \text{P}_L +  m_d \xi^A_d \text{P}_R\right)d 
+\dfrac{\sqt m_\tau}{v}H^+ \xi^A_\tau \overline{\nu}_L\tau_R^{}
+\hc
\right\},
\eea
where $\kp_f$ and $ \xi^{H,A}_f$ in type-I are 
\bea
\kp_f  = \sba+\frac{\cba}{\tb},
\quad
\xi^H_f = - \frac{\sba}{\tb} +\cba,
\quad 
\xi^A_u = - \xi^A_d =-\xi^A_\tau = \tbi.
\eea

\section{Characteristics of type-I with light charged Higgs bosons}
\label{sec:scan}

\subsection{Theoretical and experimental constraints}
\label{subsec:constraints}
We study the implication of the theoretical and experimental constraints
on type-I with a light $\ch$.
Two cases for $\mch$ are considered:
\bea
\label{eq:setup}
\mch=110,~140\gev.
\eea
The other parameters are scanned over the following ranges:
\bea
\label{eq:scan:range}
\tb &&\in [2.7,50], \quad \sba \in [0.75,1], 
\\ \nn
 \mhh &&\in [130,3000]\gev,
\quad \ma \in [15,3000]\gev, \quad m_{12}^2 \in [-3000^2,3000^2]\gev^2. 
\eea
The condition of $\tb>2.7$ makes type-I consistent with the observation of $b\to s\gm$~\cite{Haber:2015pua,Arbey:2017gmh}.
The range of $\sba$ is conservatively taken by considering the current Higgs precision data~\cite{Aad:2019mbh,CMS:2020xwi,ATLAS:2021vrm}:
the most updated results on the coupling modifiers are $\kp_Z>0.86$
and $\kp_W>0.94$ with $\kp_{W,Z} \leq 1$ at 95\% C.L.~\cite{Aad:2019mbh}. 
For $\mhh$, we avoid the case where $\mhh$ is too close to the observed Higgs boson mass.

With the prepared random parameter sets, we cumulatively impose the following constraints:
\begin{description}
\item[Step-(i) Theory+EWPD+FCNC:]
We require the parameter set to satisfy the conditions in three categories.
	\bit
	\item \underline{Theoretical constraints}
		\ben
		\item Higgs potential being bounded from below~\cite{Ivanov:2006yq};
		\item Perturbative unitarity of the amplitudes of scalar-scalar, scalar-vector, and vector-vector scatterings at high energies~\cite{Kanemura:1993hm,Akeroyd:2000wc};
		\item Perturbativity of the quartic couplings~\cite{Branco:2011iw,Chang:2015goa};
		\item Vacuum stability~\cite{Deshpande:1977rw,Barroso:2013awa}.
		\een
	The detailed expressions are referred to the references. 	
      \item \underline{Electroweak precision data}\\
      We calculate the Peskin-Takeuchi electroweak
        oblique parameters in the 2HDM~\cite{He:2001tp,Grimus:2008nb}
        and require $ \chi^2 < 7.815$ for the current best-fit results of~\cite{Zyla:2020zbs}
	\bea
	\label{eq:STU}
	S &=& -0.01 \pm 0.10,
	\quad
	T = 0.03 \pm 0.12, \quad
	U=0.02 \pm 0.11, 
	\eea
	where the correlations among the oblique parameters have been properly taken into account.%
	\item \underline{$\mathbf{b\to s\gm}$ constraints} \\
	We consider the most sensitive FCNC process
to the 2HDM,  $b\to s\gm$~\cite{Misiak:2020vlo}.
	\eit
\item[Step-(ii) Higgs precision data:] To check the consistency with the Higgs precision
  data, we use \textsc{HiggsSignals}-v2.2.0~\cite{Bechtle:2020uwn}, which yields
  the $\chi^2$ output for 107 Higgs observables~\cite{Aaboud:2018gay,Aaboud:2018jqu,Aaboud:2018pen,Aad:2020mkp,Sirunyan:2018mvw,Sirunyan:2018hbu,CMS:2019chr,CMS:2019kqw}.
Since there are five model parameters with the given $\mch$,
the number of degrees of freedom is 102.
We demand that the $p$-value be larger than 0.05.
In addition, the total width of the Higgs boson is required to be within the
experimental upper bound at 95\% C.L., i.e.,  
$\Gm_h^\tot< 9.16\mev$~\cite{Sirunyan:2019twz}.
\item [Step-(iii) Direct searches:]
Using \textsc{HiggsBounds-5}~\cite{Bechtle:2020pkv},
we calculate $r_{95 \%}$ for each direct search process at the LEP, Tevatron, and LHC, defined by
			\bea
			\label{eq:r95}
			r_{95\%} = \frac{S_{\rm type-I}}{S_{\rm obs}^{95\%}},
			\eea
where $S_{\rm type-I}$ is the predicted cross section in the model
and $S_{\rm obs}^{95\%}$ is the observed upper bound on the cross section at the 95\% C.L.
A parameter set is excluded if $r_{95\%} > 1$.			
\end{description}

\subsection{Characteristics of surviving parameters}

We perform the random scan over the full five-dimensional parameter space
and cumulatively impose the constraints in Step-(i), Step-(ii), and Step-(iii).
First, we obtained $10^6$ parameter sets that satisfy Step-(i) for $\mch=110\gev$ and 
another $10^6$ for $\mch=140\gev$.
After applying the constraints at Step-(ii),
about 24.4\% (27.4\%) of $10^6$ parameter sets survive for $\mch=110~(140)\gev$.
Step-(iii) is most powerful in restricting the model:
only 0.22\% (1.1\%) of the parameter sets after Step-(i) are allowed  for $\mch=110~(140)\gev$.
The smoking-gun process is the LHC search for $pp\to \ttop$
followed by $t \to H^+ b \to \tau\nu+b$~\cite{ATLAS:2018gfm},
which excludes
more than 99\% of the parameter sets that passed Step-(ii).\footnote{
There exists an alternative Higgs scenario, the inverted scenario,
where the heavier \textit{CP}-even scalar $H$ is the observed Higgs boson at a mass of 125 GeV~\cite{Chang:2015goa,Bernon:2015wef,Jueid:2021avn}.
To answer whether the light charged Higgs boson is also allowed 
in this exotic setup,
we scanned the parameter ranges of $\mh \in [15, 120] \gev$, $\ma \in [15,1000] \gev$, $\sba \in [-1,1]$, 
$m_{12}^2 \in  [-20000, 20000] \gev^2$, and $\tb = [2.7,50]$ for $\mch = 110, 140\gev$.
We found that 0.56\% (3.9\%) of the parameter points survive the final Step-(iii) 
for $\mch=110\gev~(140\gev)$.
A light charged Higgs boson is also feasible in the inverted scenario.
But the phenomenological signatures in the inverted scenario
are different from those in the normal scenario.
First, $h$ is lighter than the pseudoscalar $A$ in most of the viable parameter space.
Consequently the dominant decay mode of $\ch$ is $\ch \to \wpm h$ in the inverted scenario~\cite{Arhrib:2017wmo},
but $\ch \to \wpm A$ in the normal scenario.
Similarly, the decay modes for $A$ and
  $h$ are also considerably different between two scenarios.  
  Full investigation of
  the light charged Higgs in the inverted scenario warrants another study. 
}

\begin{figure}[h!]
\centering
\includegraphics[width=0.45\textwidth]{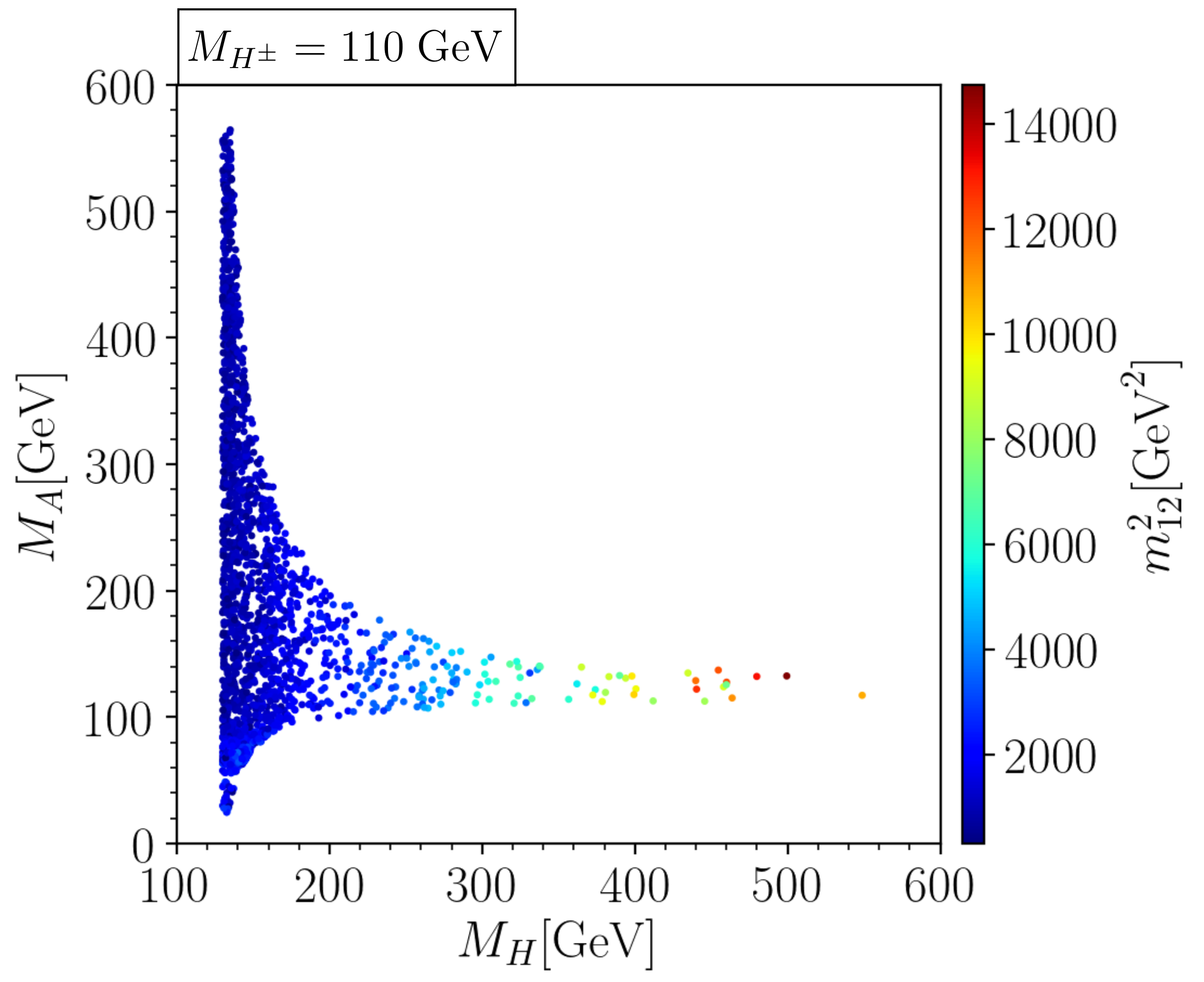}~~
\includegraphics[width=0.45\textwidth]{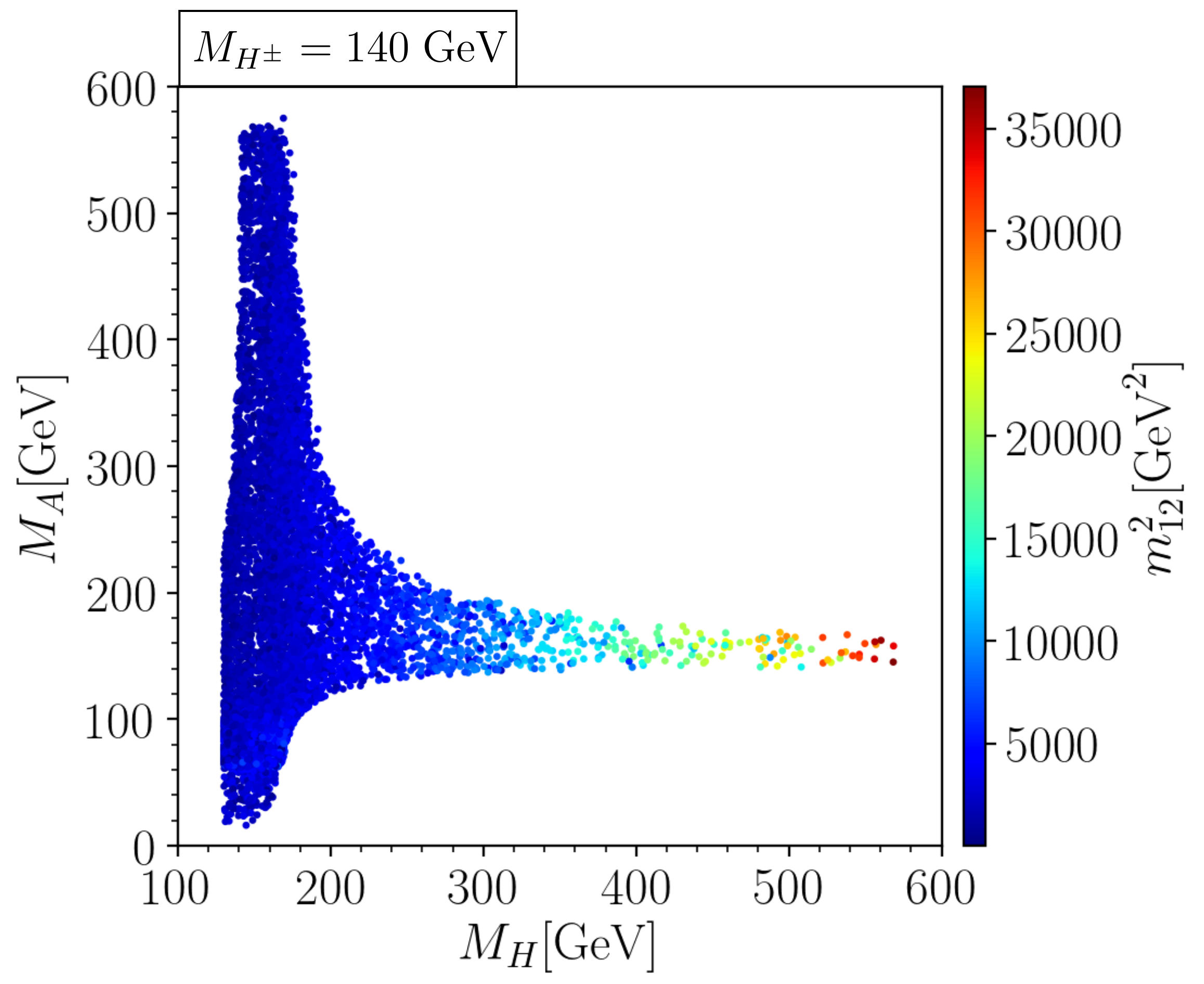}
\caption{
$\ma$ vs $\mhh$ with the color code indicating the value of $m_{12}^2$. 
We fix $\mch=110\gev$ in the left panel and $\mch=140\gev$ in the right panel. 
}
\label{fig-MHMA-m12sq}
\end{figure}

We now investigate the characteristics of the finally allowed parameter sets.
In Fig.~\ref{fig-MHMA-m12sq}, 
we show $\ma$ vs $\mhh$ with the color code indicating the value of $m_{12}^2$
for $\mch=110\gev$ (left panel) and $\mch=140\gev$ (right panel).
We first observe that $m_{12}^2$
is positive and not large:
for example, $20 \gev \lsim \sqrt{m_{12}^2} \lsim 120\gev$ when $\mch=110\gev$.
The second important aspect is that the other new scalar bosons, $H$ and $A$,
cannot be too heavy.
There exist upper bounds on their masses like $\mhh,\ma \lsim 570\gev$.
Once the charged Higgs boson is light,
partial decoupling of new Higgs bosons is not feasible in type-I.
Another intriguing feature is the correlation between $\mhh$ and $\ma$.
If $\ma$ is heavy ($\gsim 300\gev$), $\mhh$ should be light. 
If $\mhh$ is heavy above $300\gev$, 
the pseudoscalar $A$ should have an intermediate mass, 
$\ma\in [100,150]\gev$ for $\mch=110\gev$ and
$\ma\in[140,200]\gev $ for $\mch=140\gev$.
$A$ and $H$ cannot be simultaneously heavy.

Of special importance is the parameter region of $\ma< \mh/2$ 
where the exotic Higgs decay $h\to AA$ is kinematically allowed.
All the surviving parameters, consistent with the current Higgs precision data~\cite{Aaboud:2018esj,Aaboud:2018iil,Sirunyan:2018mbx,Sirunyan:2018mot,Sirunyan:2018pzn,Sirunyan:2019gou},
yield $\br(h \to AA)\lsim 10\%$.
In detail, about 50\% of the allowed parameter sets with $\ma< \mh/2$ predict $\br(h \to AA) \lsim 1\%$
while about 20\% yield $7\% \lsim \br(h \to AA) \lsim 10\%$.
The ongoing LHC searches for the exotic Higgs decay are extremely important in finding out the structure of type-I.

\begin{figure}[h!]
\centering
\includegraphics[width=0.48\textwidth]{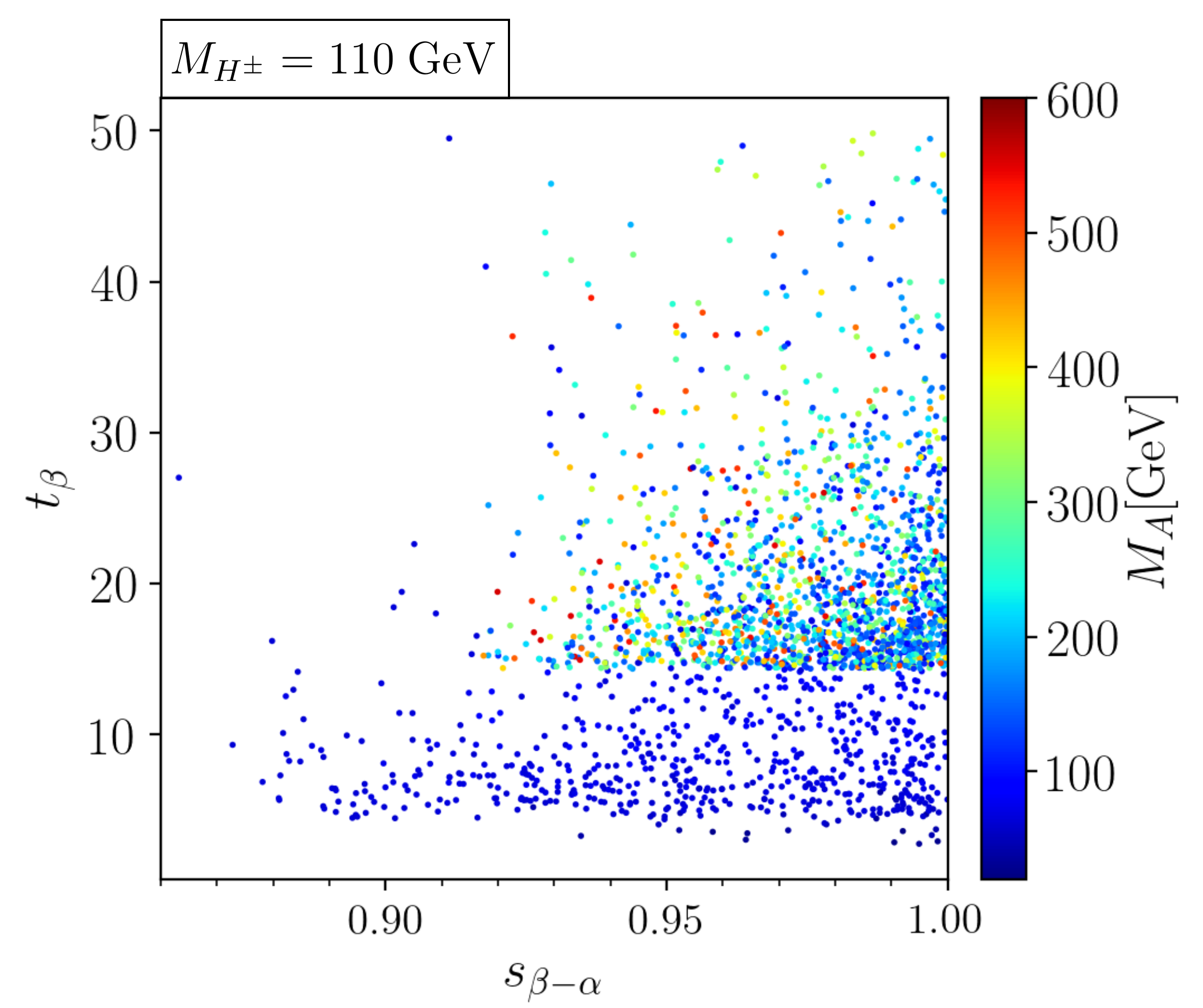}
\includegraphics[width=0.48\textwidth]{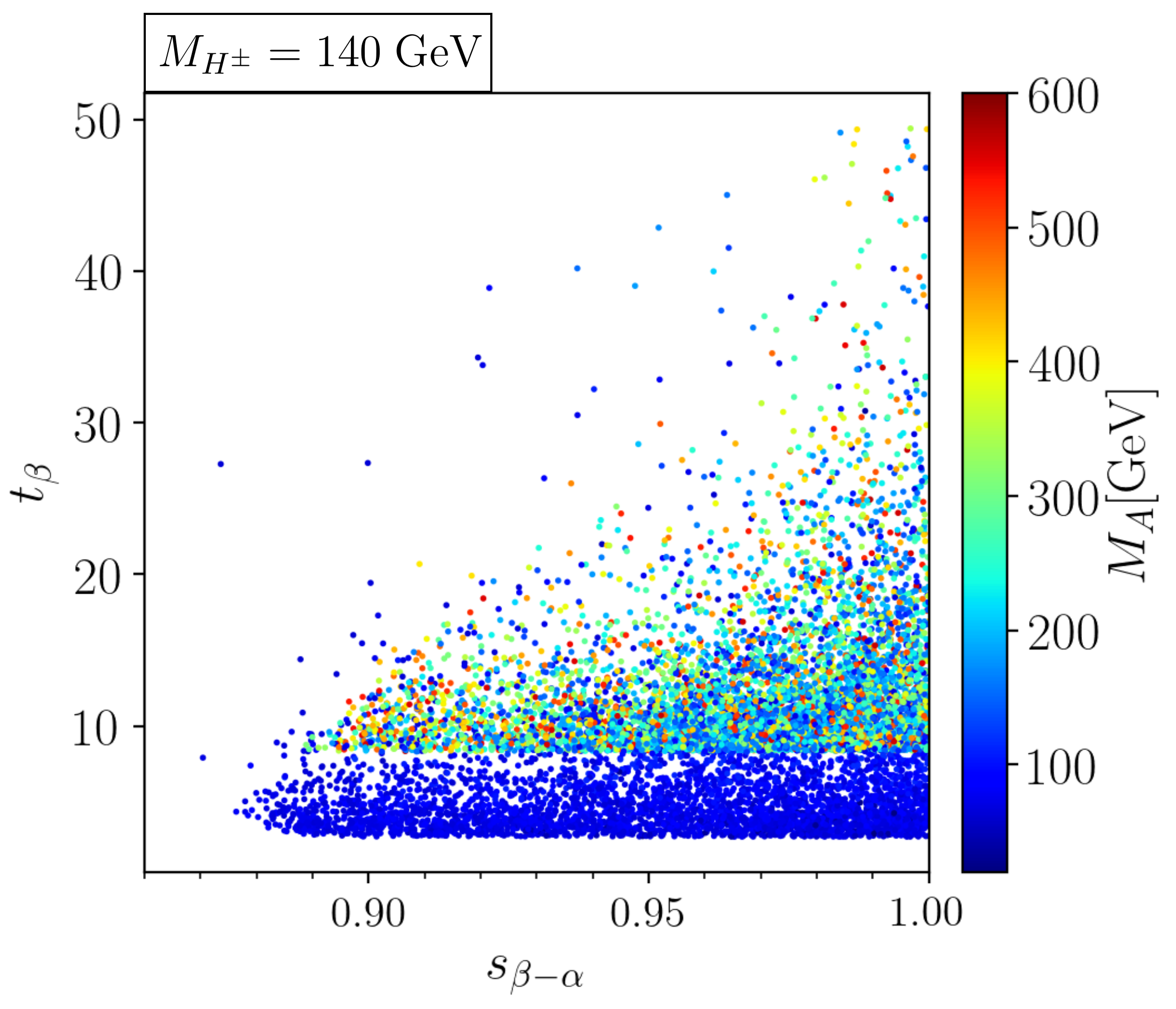}
\caption{
$\tb$ vs $\sba$ with the color code of $\ma$ for $\mch=110\gev$ (left panel) and $\mch=140\gev$ (right panel). 
}
\label{fig-sbatb-MA}
\end{figure}

Figure \ref{fig-sbatb-MA} shows $\tb$ vs $\sba$
for $\mch=110\gev$ (left panel) and $\mch=140\gev$ (right panel),
with the color code indicating $\ma$.
We observe that $\tb\lsim 10$ is still allowed, as low as $\tb \simeq 2.7$.
It seems contradictory to the
usual conclusion that no signal for the light $\ch$ at the LHC demands 
large $\tb$ in type-I.
Note that the conclusion is based on the assumption of $\mch\simeq \mhh\simeq \ma$:
the light $\ch$ decays only into the SM fermions.
Even though the mass degeneracy can easily satisfy the constraint from the Peskin-Takeuchi oblique parameters,
the current data in \eq{eq:STU} leave some room for sizable mass differences,
especially when the new Higgs bosons are not heavy.
All of the surviving points with $\tb\lsim 10$ incorporate light $\ma$,
which opens the $\ch\to A W^{\pm (*)}$ mode.
Consequently,
$\br(\ch\to \tau\nu)$ reduces
and the LHC constraint on  $\br(t \to H^+ b)\times \br(H^+ \to \tau\nu)$ can be evaded.
If $\ma$ is above the threshold of $\ch\to A W^{\pm (*)}$,
$\tb$ should be large, $\tb\gsim 15$ for $\mch=110\gev$ and $\tb\gsim 9$ for $\mch=140\gev$.
Finally, we observe that sizable deviation from the alignment limit is still possible in type-I,
like $\sba \gsim 0.87$.
As a result, the model can accommodate $H \to WW/ZZ$, $A \to h Z$, and $\ch\to W^{\pm (*)}h$.

\begin{figure}
\centering
\includegraphics[width=0.425\textwidth]{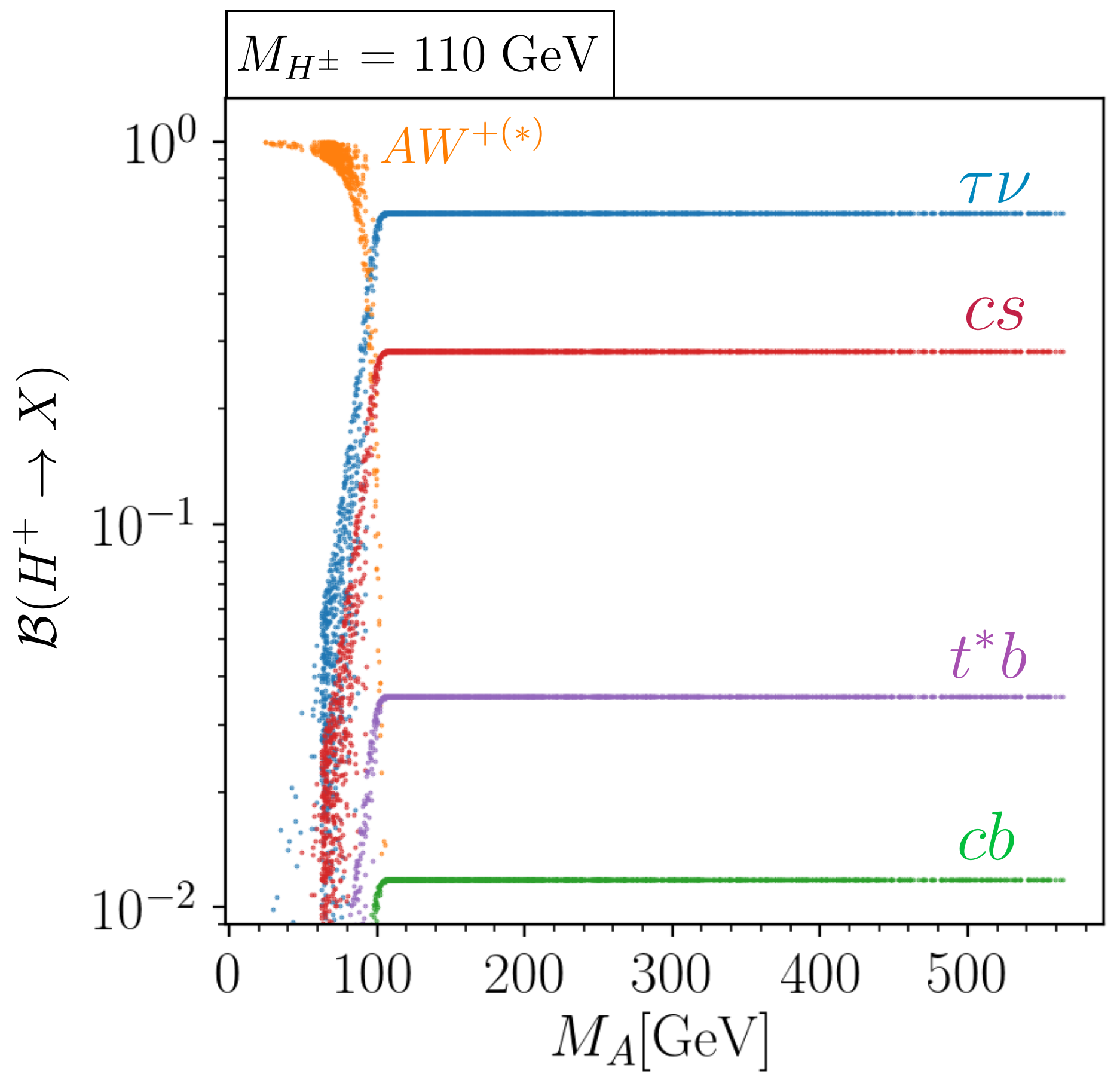}~~
\includegraphics[width=0.44\textwidth]{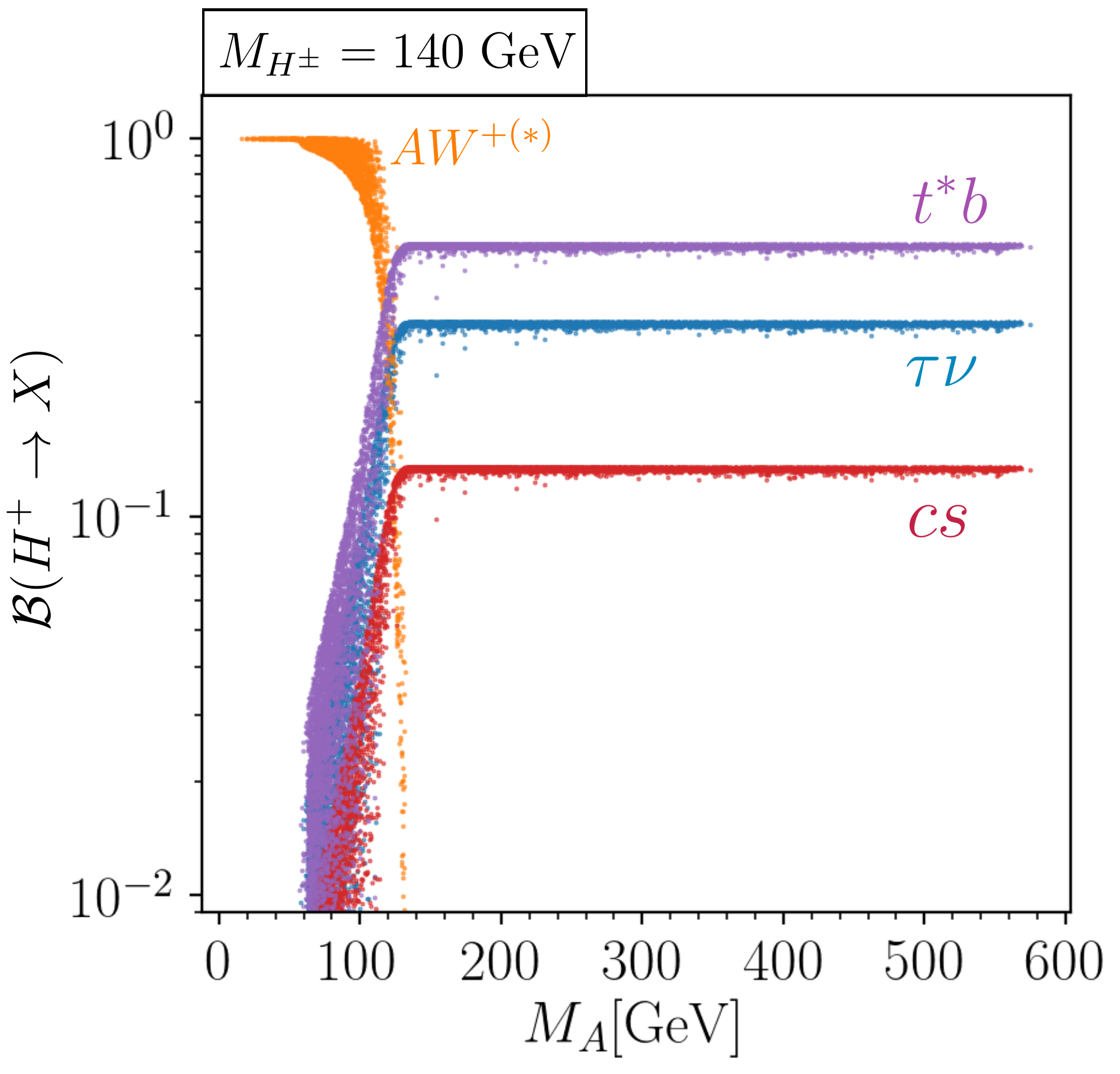}
\caption{
Branching ratios of the charged Higgs boson $\ch$ 
vs $\ma$,
predicted by all the surviving parameter points.
We fix $\mch=110\gev$ (left panel) and $\mch=140\gev$ (right panel). 
}
\label{fig-BR-cH}
\end{figure}

We move on to the next question of whether the new Higgs bosons prefer some specific
decay modes.
It is closely related to one of our goals,
a complete roadmap for the light charged Higgs boson in type-I.
The critical parameter is found to be $\ma$.
In Fig.~\ref{fig-BR-cH}, we present the branching ratios of $\ch$ vs $\ma$
for $\mch=110\gev$ (left panel) and $\mch=140\gev$ (right panel), 
where the scattered points correspond to all the surviving parameter sets.
We include the three-body decay of $A W^{\pm (*)}$.\footnote{
The three-body decay of $\ch \to A^* \wpm$ is negligible
since the Yukawa couplings of $A$ to $f\bar{f}$ are much smaller than the
gauge couplings of the $\wpm$ boson.}
For $\ch\to q \bar{q}\,'$,
we incorporate QCD radiative corrections at order $\alpha^2_s$
in the \textoverline{MS} scheme~\cite{Braaten:1980yq,Drees:1990dq,Gorishnii:1990zu}
by using \textsc{2HDMC}~\cite{Eriksson:2009ws}.
For the running fermion masses in the Higgs couplings, we resum the leading logarithmic corrections to all orders with the renormalization scale
of $\mu_R=\mch$ in the \textoverline{MS} scheme.

Figure \ref{fig-BR-cH} clearly demonstrates strong correlation between $\br(\ch\to X)$
and $\ma$.
For a light $A$ below the $A\wpm$ threshold,
$\ch\to  A W^{\pm (*)}$ is dominant, which was first pointed out in Ref.~\cite{Akeroyd:1998dt}.
If the on-shell decay is possible, the branching ratio reaches almost 100\%.
The off-shell decay also has a sizable branching ratio.
Large $\br(\ch\to  A W^{\pm (*)})$ is attributed to the gauge coupling of the $\ch$-$\wmp$-$A$ vertex.
As soon as $\ma$ crosses over the kinematic threshold,
$\ch\to \tau^\pm\nu$ mode becomes important,
yielding $\br(\ch\to\tau\nu)\simeq 60~(30)\%$ for $\mch=110~(140)\gev$.
The hadronic modes such as $t^* b$ and $c s$ are also substantial.

\begin{figure}
\centering
\includegraphics[width=0.425\textwidth]{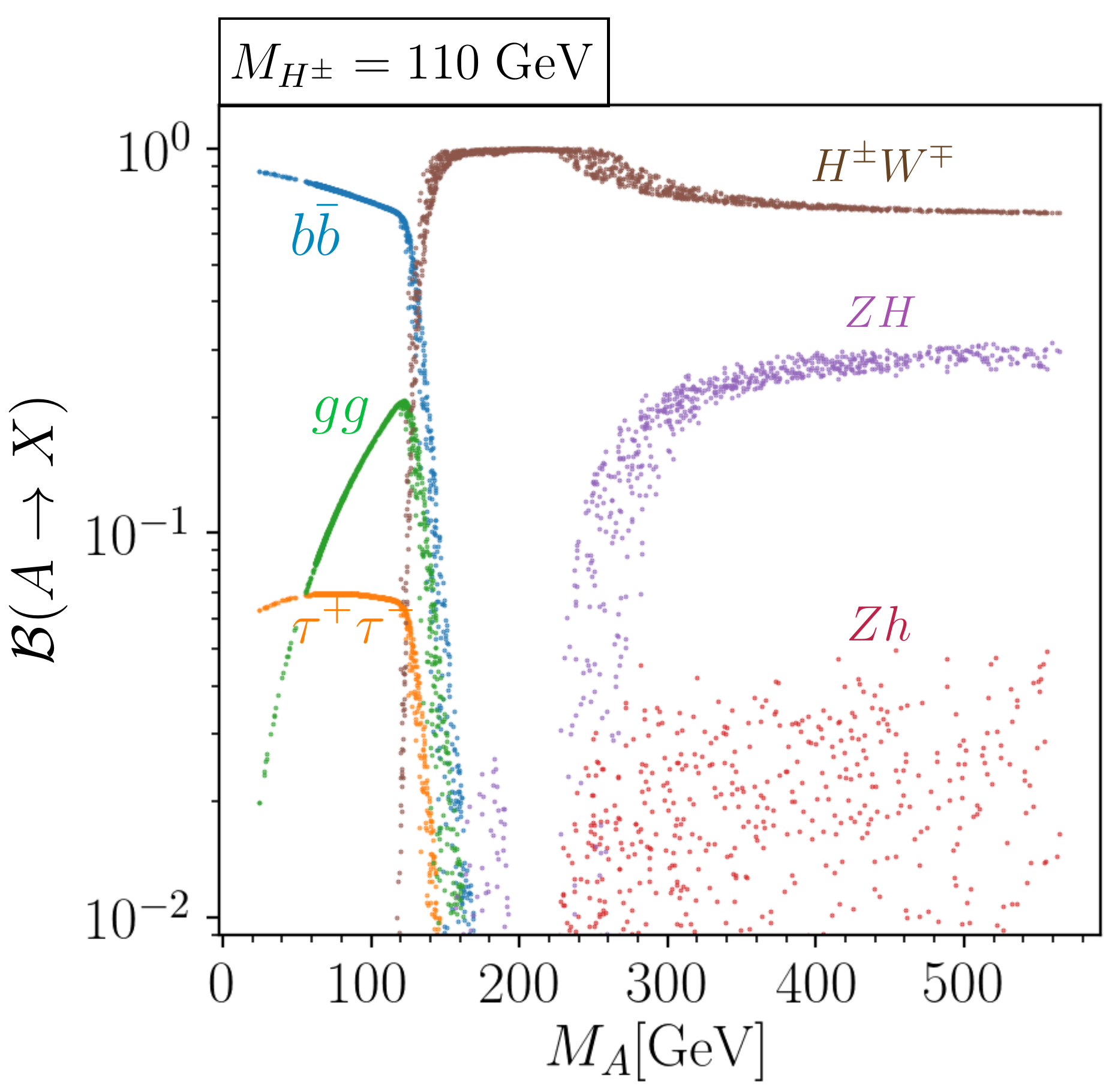}~~
\includegraphics[width=0.44\textwidth]{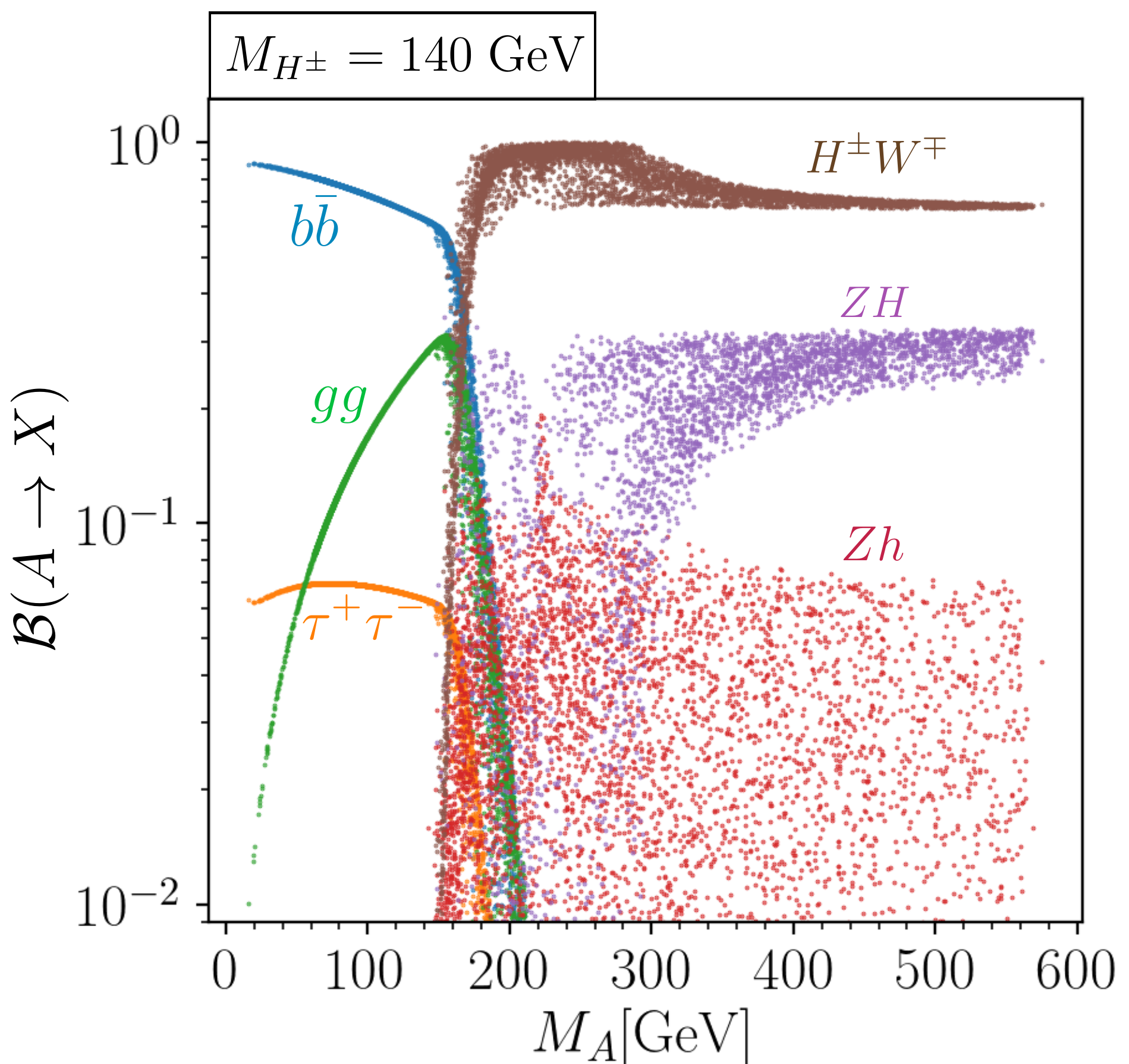}
\caption{
Branching ratios of the pseudoscalar boson $A$ vs $\ma$ for $\mch=110\gev$ (left panel) and $\mch=140\gev$ (right panel). }
\label{fig-BR-A}
\end{figure}

Figure \ref{fig-BR-A} presents the branching ratios of $A$ vs $\ma$ for
$\mch=110\gev$ (left panel) and $\mch=140\gev$ (right panel).
We see a strong correlation between $\br(A\to X)$ and $\ma$.
Below the threshold of $A\to \ch W^{\mp (*)}$ , 
$A \to \bb$ is the dominant decay mode, followed by $A\to gg$ and $A \to \ttau$.
Above the threshold, $\ch\wmp$ is the main decay mode.
Unexpected is sizable and almost constant $\br(A \to Z H)$ when $\ma\gsim 300\gev$.
The result is attributed to two factors:
the $A$-$Z$-$H$ vertex is favored by the alignment;
a heavy $\ma$ is permitted only for light $\mhh$ as shown in Fig.~\ref{fig-MHMA-m12sq}.
$\br(A\to Z h)$ is suppressed by the factor $\cba^2$.
$\br(A\to \ttop)$ is also small
because $\ma$ above the kinematic threshold ($\ma>2 m_t$)
requires large $\tb$ which suppresses the top quark Yukawa coupling to $A$.

\begin{figure}
\centering
\includegraphics[width=0.44\textwidth]{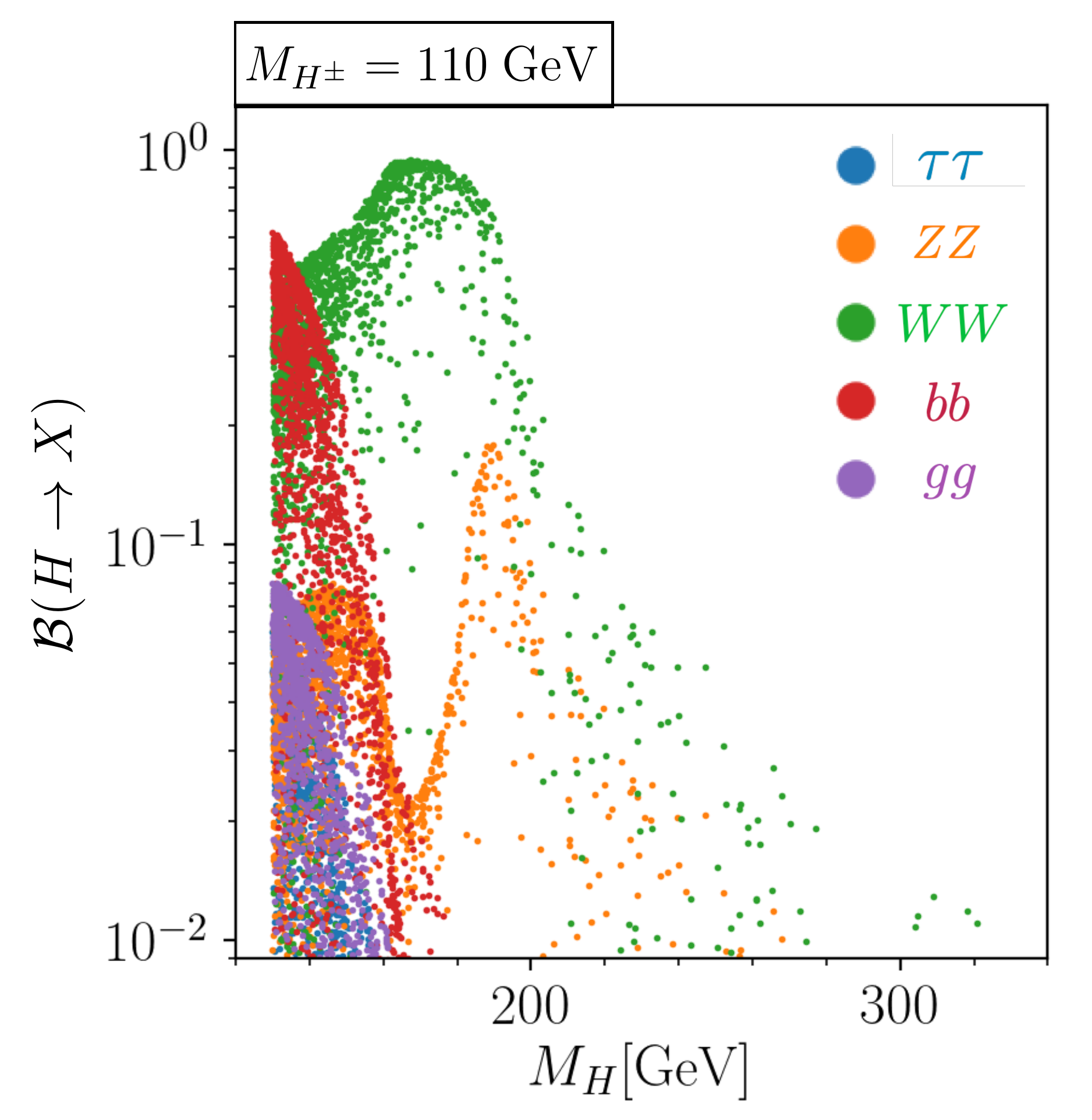}
\includegraphics[width=0.44\textwidth]{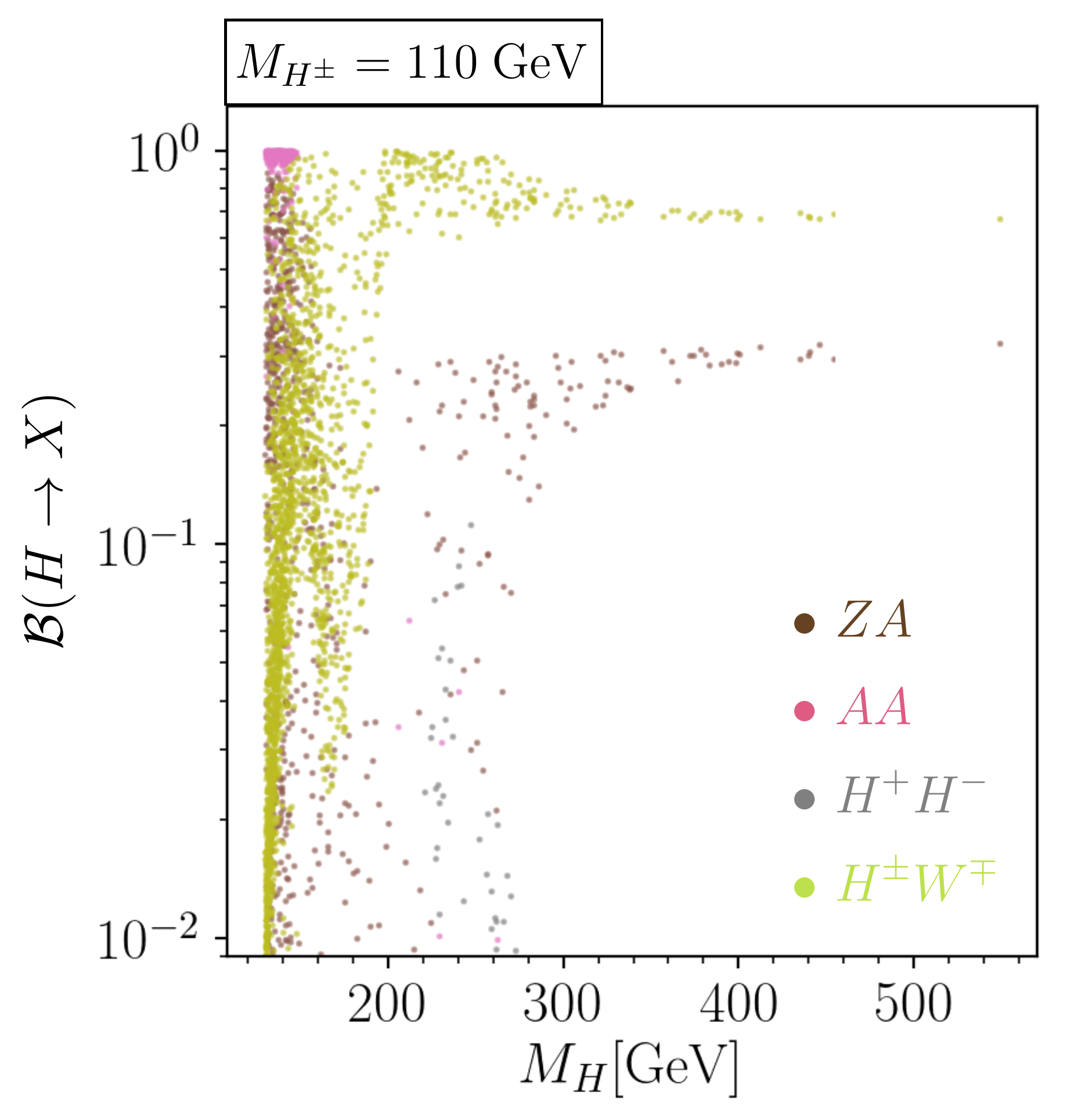}
\caption{Branching ratios of $H$ decaying into the SM particles (left panel) and into one or two new Higgs bosons (right panel)
as functions of $\mhh$. 
We fix $\mch=110\gev$.
}
\label{fig-BRH}
\end{figure}

Unlike $\ch$ and $A$, 
the heavy \textit{CP}-even $H$ shows the wide variety of decay patterns.  
In Fig.~\ref{fig-BRH}, we present $\br(H\to X)$ vs $\mhh$ 
for $\mch=110\gev$.
Nine decay modes ($\ttau$, $ZZ$, $\ww$, $\bb$, $gg$, $ZA$, $AA$, $H^+ H^-$, and $\ch\wmp$)
are all mixed up,
particularly when $\mhh < \mch+m_W$:
for a clear distinction,
we present the decays into the SM particles in the left panel and the decays into one or two new Higgs bosons
in the right panel.
The complication is from the involvement of two model parameters, $\mhh$ and $\sba$.
Another important feature is that  below the threshold of $H \to\ch\wmp$, $H \to ZZ$ and $H\to \ww$ become substantial,
which represents a deviation from the Higgs alignment limit.
Above the threshold,
$H \to \ch W^\mp$ is dominant in a large portion of the allowed parameter space
(see the right panel of Fig.~\ref{fig-BRH}).
The sizable $H$-$\ch$-$W^\mp$ vertex provides a new production channel for
the light charged Higgs boson in type-I.

\begin{figure}
\centering
\includegraphics[width=0.44\textwidth]{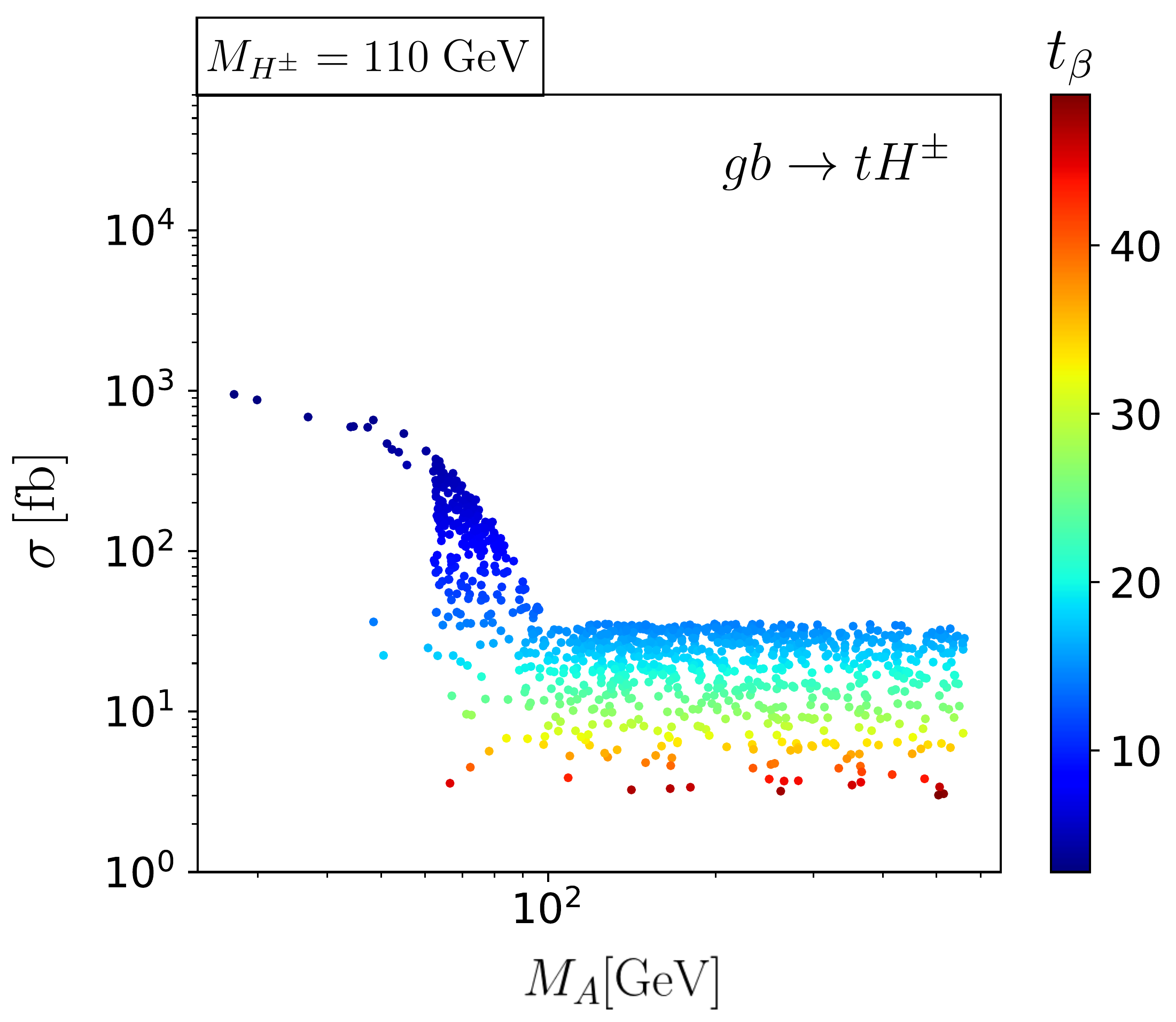}
\includegraphics[width=0.44\textwidth]{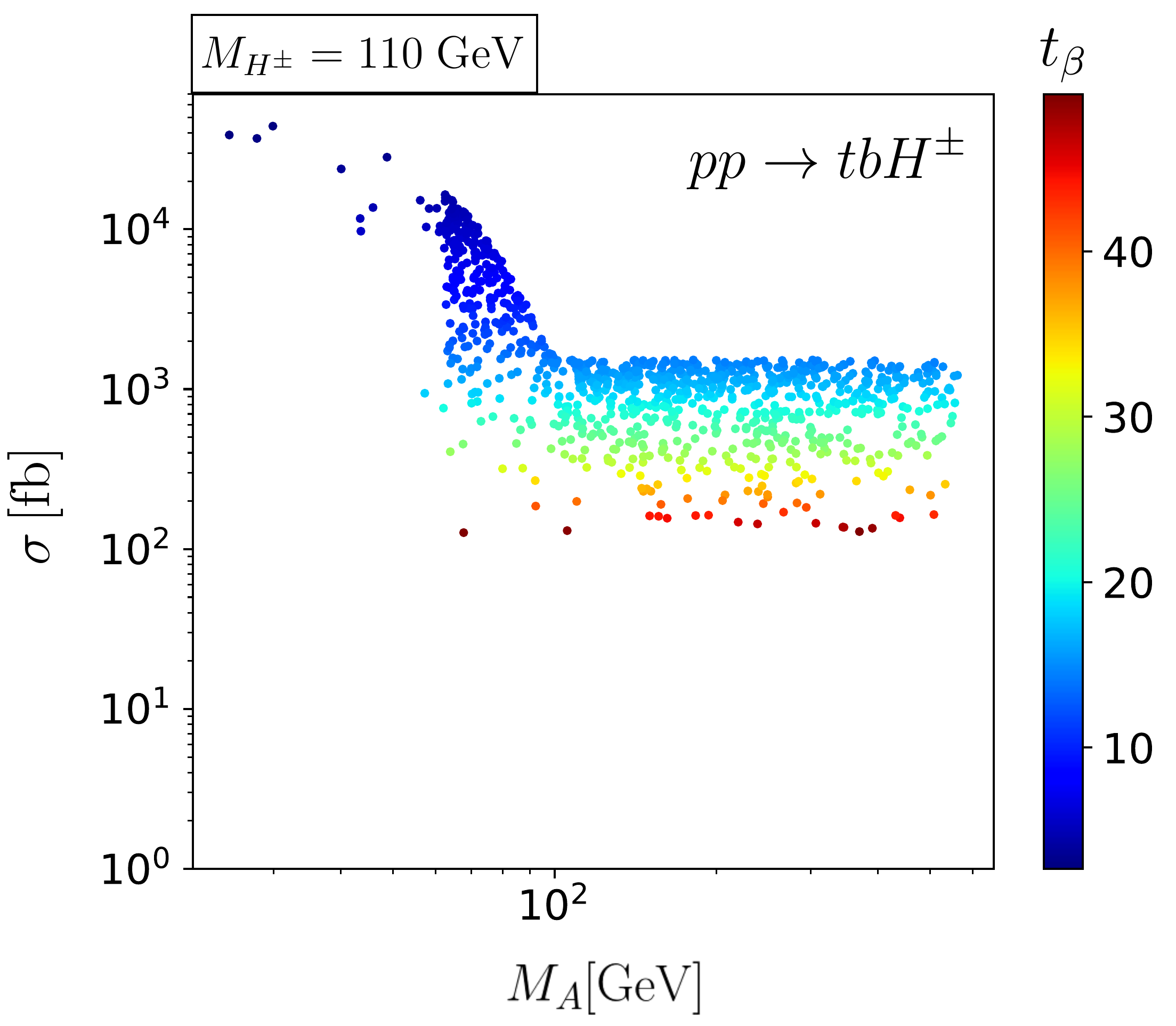}
\caption{Parton-level cross sections of $g b \to t \ch$ (left panel)
and $pp \to b t \ch$ (right panel) at the 14 TeV LHC. We fix $\mch=110\gev$
and impose $p_T^b > 30\gev$ and $|\eta_b|<2.5$.
}
\label{fig-conventional-xsec}
\end{figure}

The final study in this section is on the conventional production channels
of the light charged Higgs boson, $g b \to t \ch$ and $pp \to b t \ch$,
which resort on a \textit{single} $\ch$ production.
We calculate the parton-level cross sections at the LHC with $\sqrt{s} = 14$ TeV,
as scanning over the viable parameter space.
We used \textsc{MadGraph\_aMC@NLO}~\cite{Alwall:2014hca}
with \textsc{NNPDF31\_lo} parton distribution
function (PDF) set~\cite{Ball:2017nwa} in the five quark flavor scheme.
Figure \ref{fig-conventional-xsec} presents,
as a function of $\ma$, the cross sections of  $g b \to t \ch$ in the left panel
and those of $pp \to b t \ch$ in the right panel.
The color code indicates the value of $\tb$.
We fix $\mch=110\gev$ and demand $p_T^b > 30\gev$ and $|\eta_b|<2.5$.
For $pp \to b t \ch$,
we included not only the gluon fusion production but also $q\bar{q}$ annihilation production.
Since $\mch$ is considerably lighter than the top quark mass,
the cross section of $pp \to b t \ch$, mainly through the top quark pair production followed by $t\to b\ch$,
is much larger than that of $g b \to t \ch$.
For the given $\ma$, which governs the decays of $\ch$ and $A$,
the cross sections of two production channels show wide varieties.
Instead, the value of $\tb$ strongly correlates with the cross sections,
which are proportional to  $1/\tb^2$.
So the conventional production channels,
which are more model-dependent,
complicate the search for the light charged Higgs boson. 
Furthermore, these processes of a single $\ch$ production suffer from huge backgrounds.
In this work, therefore, we consider the unconventional production channels of the light charged Higgs boson
through the pair production, 
avoiding the direct decay from the top quark.

\section{Production of light charged Higgs bosons at the LHC}
\label{sec:production}
Based on the characteristics of the viable parameter space, 
we develop the search strategies for the light $\ch$ in type-I.
Since $\ma$ is shown to be the key parameter,
we divide the parameter space into two regions,
the light $A$ case and the heavy $A$ case
with the threshold of $\ma^{\rm threshold} \simeq 100~(120)\gev$ for $\mch=110~(140)\gev$.
When $A$ is light, $\ch$ dominantly decays into $A \wpm$, and $A$ decays into $\bb$.
In the heavy $A$ case,
$\ch\to\tau\nu$ and $A \to \ch\wmp$ are main decay modes.
For the decays of $H$,
we focus on $H \to  \ch\wmp$
to find new production channels of the light charged Higgs boson at the LHC.

\begin{table}
\begin{tabular}{|c||c|c|}
\hline
& light $A$ case &  heavy $A$ case \\ \hline\hline
\multirow{3}{*}{Target decay modes}&  $\ch \to A W^{\pm(*)}$ & $\ch\to \tau\nu$ \\ \cline{2-3}
  &  $A \to \bb$ &   $A \to \ch W^{\mp(*)}$ \\ \cline{2-3}
  &  \multicolumn{2}{c|}{$H \to \ch \wmp$ } \\ \hline \hline
{Initial production} & \multicolumn{2}{c|}{Final states} \\ \hline\hline
$gg\to  h/H/A \to \ch W^\mp$ & $[\bb\wpm]\wmp $ &  $[\tau\nu]\wpm$  \\ \hline
$q\bar{q}'\to W^* \to  \ch h$ &  $[\bb \wpm] \,h$ & $[\tau\nu] h$  \\ \hline
$gg\to H \to A Z$   & $\bb Z$ & $[\tau\nu] \wpm Z$ \\ \hline
$gg\to H Z,~\qq\to Z^*\to H Z$  & $[\bb \wpm] \wmp Z$ & $[\tau\nu] \wpm Z$ \\ \hline
$q\bar{q}'\to W^* \to \ch A$ &  $[\bb \wpm] \bb$ & $[\tau\nu] [\tau\nu] \wpm$ \\ \hline \hline
$q\bar{q}'\to W^* \to \ch H$ &  ~~~$[\bb W] [\bb W]\wpm$\checkmark ~~~& $[\tau\nu][ \tau\nu] \wpm$ \\ \hline
$pp \to H^+ H^-$  &  $[\bb \wpm][\bb \wmp]$  \checkmark & $[\tau\nu][\tau\nu]$ \checkmark  \\ \hline
$\qq \to Z^* \to H A$  & $[\bb\wpm]\bb\wmp$ & ~~~$[\tau\nu][\tau\nu]\wpm\wpm$  \checkmark~~~ \\ \hline
$gg \to HH$ & ~~~$[\bb W][\bb W]WW$~~~ & ~~~$[\tau\nu][\tau\nu]\wpm\wpm$  \checkmark~~~ \\ \hline
$gg \to AA$ &  $\bb\bb$ & ~~~$[\tau\nu][\tau\nu]\wpm\wpm$  \checkmark~~~ \\ \hline
\end{tabular}
\caption{\label{table:production}
For the light and heavy $A$ cases,
the production channels of one or two charged Higgs bosons at the LHC, 
and the subsequent final states from
the targeted decay modes of $\ch$, $A$, and $H$. The particles inside a square
bracket in the final states are from the decay of one charged Higgs boson. 
The processes with a checkmark are expected to have high LHC discovery potential.}
\end{table}

In Table \ref{table:production},
we summarize the possible production channels of the
light $\ch$  at the LHC,
and the final states from the targeted decay modes of $\ch$, $A$, and $H$. 
To emphasize the decay products of a charged Higgs boson,
we adopt the notation of a square bracket:
$[ijk]$ denotes $\ch\to ijk$.
To find the processes with high LHC discovery potential,
we focus on the production of \emph{two} charged Higgs bosons,
which is more challenging for the background to mimic.
We also consider the process with additional tagging particles that help to tame the background
and increase the significance.
And we avoid the signal processes with too small cross section, below about $1\fb$.
In Table \ref{table:production}  we put the checkmarks 
on the candidate processes.

In this regard, we study the following four channels:
\bit
\item For the light $A$ case,
	\bit
	\item $[bb  W][bb W]$:\\
	The signal cross section is
	\bea
	\sg_{[bb  W][bb W] }&=& 
	\left[\sg (\qq\to H^+ H^-) + \sg(gg\to H^+ H^-) \right]
	\\ \nn &&\times \br(H^+ \to A W^+)^2 \times \br(A \to \bb)^2.
	\eea
	\item $[bb W][bb W]W$:\\
	The total signal rate is
	\bea
	\sg_{[bb W][bb W]W}
	&=&
	\left[ \sg(\qq' \to W^{+ *} \to H^+ H)	+\sg(\qq' \to W^{- *} \to H^- H)	\right]
	\\ \nn && 
	\times 2 \br(H \to H^+ W^-) \br(H^+\to A W^+)^2 \br(A \to \bb)^2.
	\eea
	Four different charge conjugation combinations are to be summed.
	\eit
\item For the heavy $A$ case,
	\bit
	\item $[\tau\nu][\tau\nu](j)$:\\
	The signal cross section is
	\bea
	\sg_{[\tau\nu][\tau\nu]j} = \left[ \sg (pp \to H^+ H^-) +  \sg (pp \to H^+ H^- j)\right]
	\times \br(\ch\to \tau\nu)^2.
	\eea
	A pair of charged Higgs bosons is produced at the LHC via
	the Drell-Yan process and the gluon fusion.
	\item $[\tau\nu][\tau\nu]WW$:\\
	We have
	\bea
	\label{eq:taunutaunuWW}
	\sg_{[\tau\nu][\tau\nu]WW} &=& 	
	\left[ \sg(\qq\to Z^* \to HA) \times 4 \,\br(H \to H^+ W^-)\br(A \to H^+ W^-) \right.
	\\ \nn && \, + \, \sg (gg \to HH) \times 4 \,\br(H \to H^+ W^-)^2  
	\\ \nn && \, + \,  \sg( gg \to AA)\times 4 \,\br(A \to H^+ W^-)^2	\left. \right]
	\times \, \br(H^+ \to \tau \nu)^2,
	\eea
	where the factor of four covers four different combinations of charge conjugation.
	Half of them correspond to the same-sign $W$'s, 
	$\tau^+ \tau^+ W^- W^- \nu\nu$ and $\tau^- \tau^- W^+ W^+ \nu\nu$.
	\eit
\eit

\begin{figure}
\centering
\includegraphics[width=0.44\textwidth]{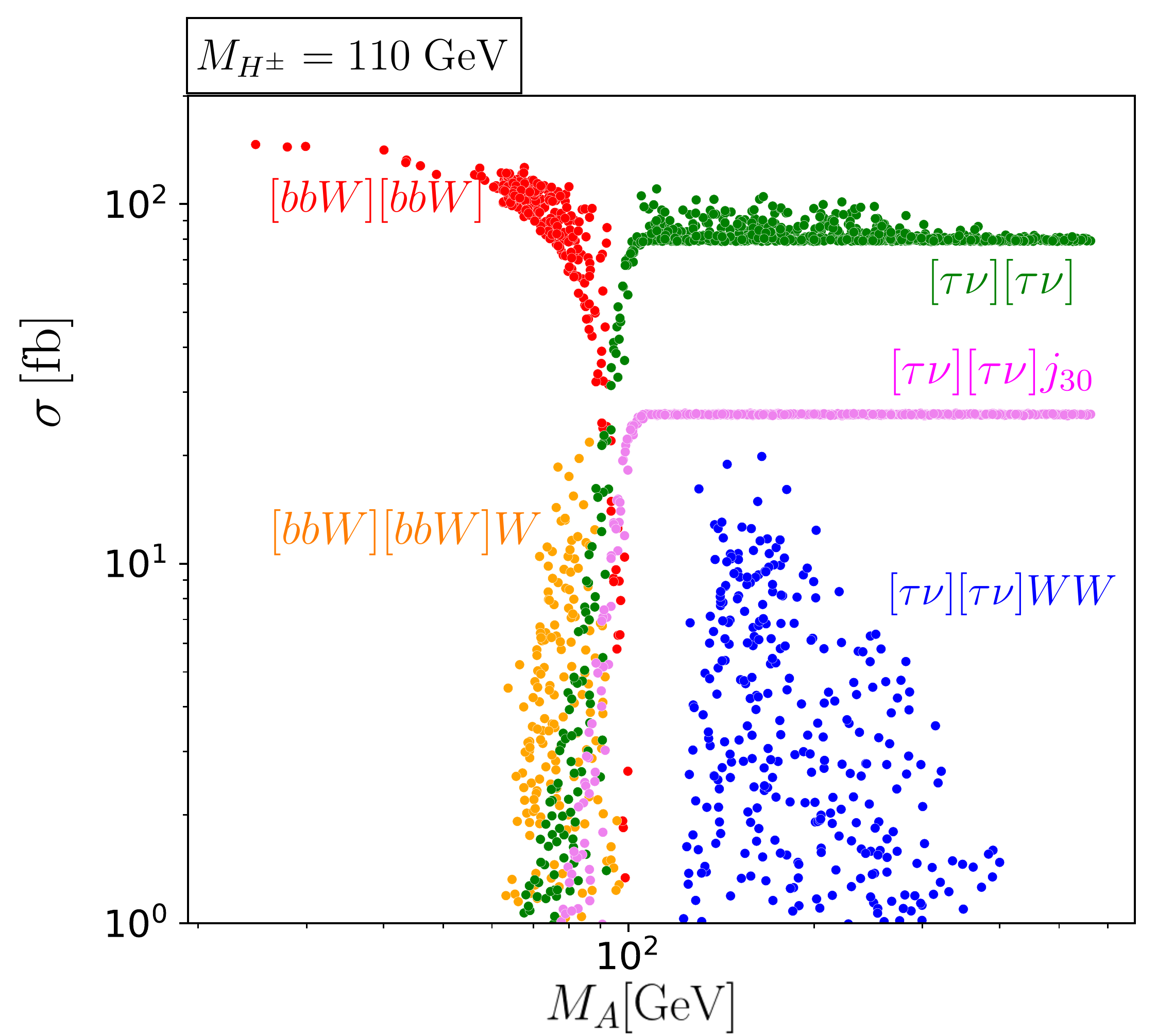}~~
\includegraphics[width=0.44\textwidth]{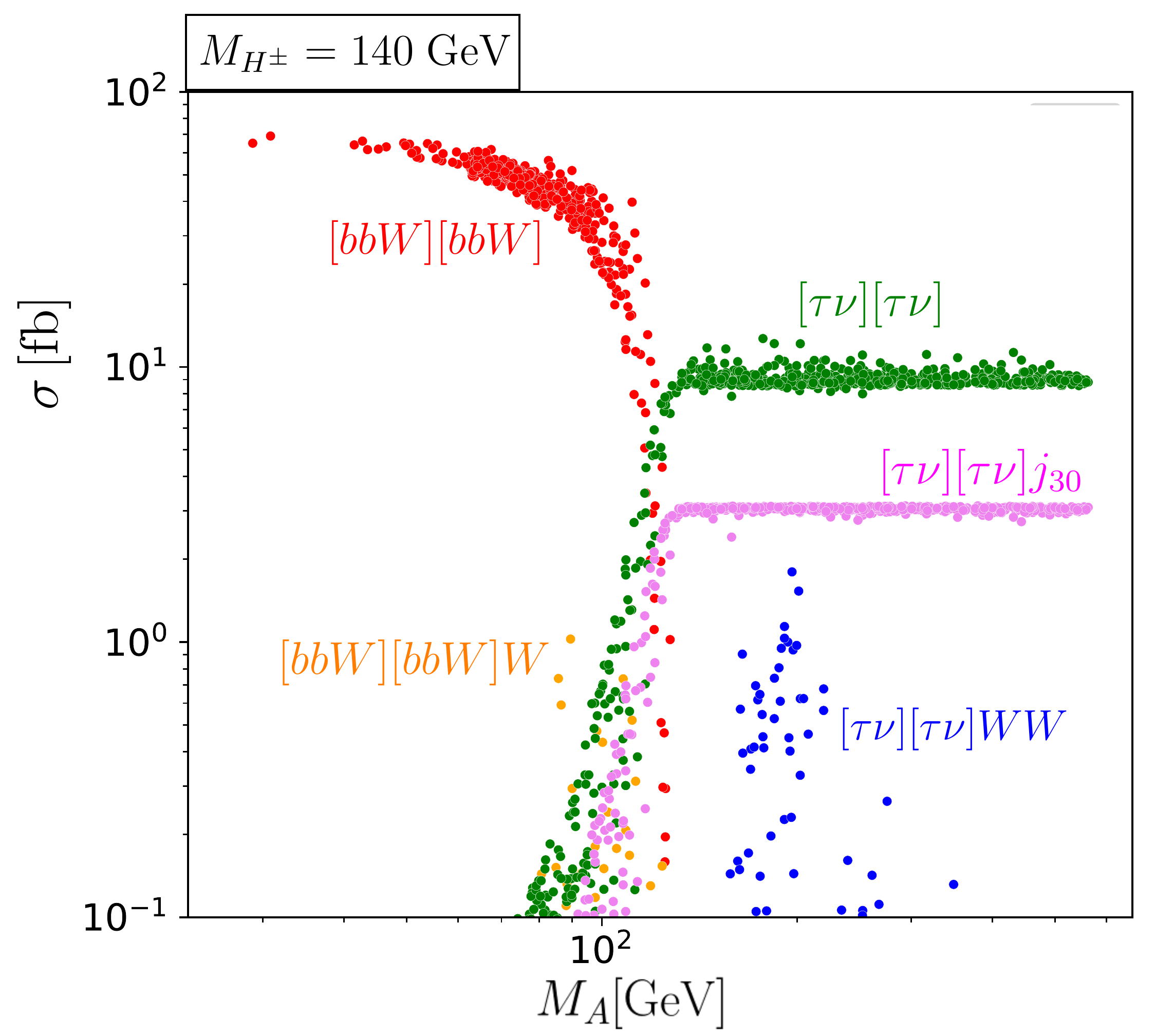}
\caption{
Parton-level cross sections of the signal in the final states of
$[\bb  W^+]\;[\bb W^-]$, $[\bb W][\bb W]W$, $[\tau\nu]\;[\tau\nu]$, $[\tau\nu]\;[\tau\nu]j_{30}$, and
$[\tau\nu]\,[\tau\nu]WW$ at the 14 TeV LHC.
The particles inside a square bracket represent the decay products of a charged Higgs boson,
and $j_{30}$ denotes a jet with $p_T^j >30\gev$.
We consider $\mch=110\gev$ (left panel) and $\mch=140\gev$ (right panel). }
\label{fig-production}
\end{figure}

Over the whole parameter space that satisfies all  the theoretical and
experimental constraints at Step-(i), Step-(ii), and Step-(iii),
we calculate the parton-level cross sections at the LHC with $\sqrt{s} = 14$ TeV.
We use \textsc{MadGraph\_aMC@NLO}~\cite{Alwall:2014hca}
with \textsc{NNPDF31\_lo} parton distribution
function (PDF) set~\cite{Ball:2017nwa}.
The renormalization and factorization scales are set to
$\mu_R =\mu_F =  \sum_i (1/2) \sqrt{p_{T,i}^2 + m_i^2}$.
Since the 2HDM \textsc{UFO} file in the \textsc{MadGraph} misses some important decay modes of new scalar bosons
such as $\ch\to cs$ and $A \to gg$,
we modified the values of the  extra scalar decay widths 
in the \textsc{MadGraph} input cards to match the output of the \textsc{2HDMC}~\cite{Eriksson:2009ws}.
In Fig.~\ref{fig-production}, we present the parton-level cross sections
for $\mch=110\gev$ (left panel) and $\mch=140\gev$ (right panel).
Here $[\tau\nu][\tau\nu]j_{30}$ denotes the pair production of charged Higgs bosons
with one extra jet from initial state radiation (ISR).
The subscript in $j_{30}$ points out the additional requirement of $p_T^j >30\gev$.
As shown below, including an extra-jet emission considerably improves the
signal significance.

Figure \ref{fig-production} clearly demonstrates the crucial role of $\ma$
in the LHC phenomenology of the light $\ch$ in type-I.
For $\ma<\ma^{\rm threshold}$,
only the process $pp\to H^+ H^- \to [bbW][bbW]$ (red points) has sizable cross sections,
which reaches about $100\fb$ for $\mch=110\gev$ and about $80\fb$ for $\mch = 140\gev$.
An advantage of this process is that the cross sections have small variations over
all the surviving parameters.
There are two reasons.
First, the main production of a charged Higgs boson pair, the Drell-Yan process,
is determined solely by  $\mch$.
Second, the decays of $\ch \to A\wpm$ and $A\to \bb$ are dominant for light $\ma$,
irrespective to $\tb$
(see Figs.~\ref{fig-BR-cH} and \ref{fig-BR-A}).

When the light $\ma$ approaches $\ma^{\rm threshold}$,
the signal rate of $[bbW][bbW]\wpm$ (yellow points) can be substantial 
when various conditions fit exquisitely.
The parameters with $\ma \simeq \ma^{\rm threshold}$ strongly prefer heavy $\mhh$: see Fig.~\ref{fig-MHMA-m12sq}.
Then a large portion of the parameter space yields sizable branching ratio for $H \to \ch\wmp$.
The production of $q \bar{q}' \to W^* \to \ch H$,
favored by the Higgs alignment,
is followed by $H \to \ch\wmp$ and $\ch \to A\wpm$.
The final state becomes $[bbW][bbW]\wpm$.

As soon as $\ma$ exceeds $\ma^{\rm threshold}$,
the cross section of $[bbW][bbW]$ rapidly drops and
$pp\to H^+ H^- \to [\tau\nu][\tau\nu]$ becomes dominant.
The cross section of $[\tau\nu][\tau\nu]$ is almost constant 
because $\br(\ch\to\tau\nu)$ is nearly constant for heavy $\ma$.
We also show the signal rate of $[\tau\nu][\tau\nu] j_{30}$ (magenta points).
Although it is a $2\to 3$ QCD process,
$g q \to H^+ H^- q$ is benefited by the high gluon luminosity.
The extra-jet emission is known to be useful 
in improving the significance, particularly for rare NP processes~\cite{Goncalves:2018qas}.
Furthermore, it provides more kinematic control to suppress the backgrounds.

Finally, we exhibit the cross sections of $[\tau\nu][\tau\nu]WW$ (blue points),
which become sizable
for moderately heavy $\ma$, above $\ma^{\rm threshold}$ but below about $250\gev$.
The pseudoscalar mass in this range demands $\mhh$ above the threshold of $H \to\ch\wmp$,
as shown in Fig.~\ref{fig-MHMA-m12sq}.
As a result, both $A$ and $H$ decay into $\ch\wmp$ with a non-negligible branching ratio.
The associated production of $H$ and $A$ mediated by $Z$,
which is preferred by the Higgs alignment limit,
leads to $[\tau\nu][\tau\nu]WW$.
Note that the gluon-fusion productions of $HA$, $HH$, and $AA$ also generate the same final state.

\section{Signal-background analysis for $[bbW][bbW]$, $[\tau\nu][\tau\nu]$, and
  $[\tau\nu][\tau\nu]WW$}
\label{sec:results}

\begin{table}
{\footnotesize
\begin{tabular}{|c|c|c|c|}
\hline
Signal & \multicolumn{2}{c|}{Benchmark point} & Backgrounds \\ \hline\hline
 $[\tau\nu][\tau\nu]$ & BP--1 & $\mch=110\gev$, $\mhh=138.6\gev$, $\ma=120.7\gev$ & $Wjj$, $Zjj$, $\ttop jj$ \\ 
 & & $\tb=16.8$, $\sba=0.975$, $m_{12}^2 =1089.7\gev^2$ & $WWjj$, $WZjj$, $ZZjj$ \\ \hline
$[\tau\nu][\tau\nu]W_{\ell^\pm \nu} W_{\ell^\pm \nu}$   & BP--2 & $\mch=110\gev$, $\mhh=138\gev$, $\ma=145\gev$ & 
$W^+ W^- W^-$, $W^- W^+ W^+$  \\ 
& & $\tb=18$, $\sba=0.999$, $m_{12}^2 =1043\gev^2$ & $\ttop \wpm$, $\ttop Z$, $\hsm Z$, $ZZ$\\ 
\hline
 $[bbW_{\ell \nu}][bb W_{qq'}]$ & ~BP--3~ & $\mch=110\gev$, $\mhh=134\gev$, $\ma=29\gev$ & $WW$, $ZZ$, $ZZ \bb$ \\ 
 & & $\tb=3.9$, $\sba=0.967$, $m_{12}^2 =533\gev^2$ & $\ttop$, $t V$, $\hsm V$, $\ttop \hsm/V$ \\ \cline{2-4}
\hline
\end{tabular} 
}
\caption{\label{table:BP:backgrounds}
Benchmark points for three target processes of a light charged Higgs boson at the HL-LHC.
The main backgrounds are also listed, with $V=\wpm,Z$.
}
\end{table}

In the previous section,
we calculated the parton-level cross sections of the proposed channels
to probe the light $\ch$ in type-I.
Although their magnitudes are not small, 
the discovery potential depends on how efficiently we isolate the signal from the overwhelming backgrounds. 
In this section, we develop the search strategies for a fully-fledged signal-to-background optimization
which relies upon sophisticated tools that include hard-scattering matrix elements, resonance decays, parton showers, hadronization, hadron decays, and a simplified detector's response.
Targeting the HL-LHC,
  we perform  detailed studies of the following three processes:
$[\tau\nu][\tau\nu](j)$, $[\tau^\pm\nu][\tau^\pm \nu]W^\mp W^\mp$, and $[bbW][bbW]$. 
For each channel,
we adopt the benchmark set in Table \ref{table:BP:backgrounds}.
We also list the backgrounds. 
The benchmark points
for $[\tau\nu][\tau\nu]$ and $[\tau\nu][\tau\nu]WW$
are representative of the process because all the allowed parameters
yield similar signal rates.
But the benchmark point for $[bbW][bbW]$
is chosen to maximize the signal rate.

Before getting into the detailed analysis for each process,
we present the common ingredients.
For the Monte Carlo event generation of the signal and backgrounds,
we use the 2HDM \textsc{UFO} file~\cite{Degrande:2011ua} and \textsc{MadGraph\_aMC@NLO} version 2.6.7.~\cite{Alwall:2014hca}
with the \textsc{NNPDF31\_lo} set of parton distribution functions~\cite{Ball:2017nwa}.
As in the previous section,
the input cards in the \textsc{MadGraph\_aMC@NLO} are modified in accordance with 
the values of 2HDMC~\cite{Eriksson:2009ws}.
We use the default settings in the run-card of \textsc{Madgraph5}
such as $p_T>10\gev$, $|\eta| < 2.5$, and $\Delta R(\ell,\ell) \geq 0.4$,
where $\Delta R = \sqrt{(\Delta \eta)^2 + (\Delta \phi)^2} $.
The resulting parton-level events are passed to \textsc{Pythia} version 8.243 to add parton showering, hadronization, and hadron decays~\cite{Sjostrand:2007gs}.
We perform a fast detector simulation of the signal and 
backgrounds using the \textsc{Delphes} version 3.4.2~\cite{deFavereau:2013fsa}. 
Jet is clustered according to the anti-$k_T$ algorithm~\cite{Cacciari:2011ma} with a
jet radius $R = 0.4$.
Since we demand to trigger at least one charged lepton, 
we do not include the pileup effects. 
We also turn off the multiple parton interactions from the soft QCD contribution
at the level of \textsc{Pythia~8}.
Under the above setup, we generate the signal and background events, which are to be called ``Initial events" in what follows.

We now turn into the discussion of the object identification,
which consists of $\tau$--tagging, $b$--tagging, and a charged lepton.  
The quality of $\tau$--tagging  is crucial and vital for $[\tau\nu][\tau\nu]$ and
$[\tau\nu][\tau\nu]WW$.
A tau lepton that decays hadronically, denoted by $\tau_{\rm h}$ in what follows, 
can be distinguished from a QCD jet by fewer particle multiplicity
and more localized energy deposits.
Recently, the $\tau_{\rm h}$--tagging efficiency has increased significantly
with the improvements in $\pi^0$ reconstruction and multivariate
discriminants~\cite{CMS:2018jrd}.
At the \textsc{Delphes} level,
we set the $\tau_{\rm h}$--tagging efficiencies and the mistagging  rates of a light jet ($j$) or
the $b$ jet as $\tau_{\rm h}$:
\footnote{The CMS collaboration has  measured
the misidentification probability of a $b$ jet as
$\tauh$ by using the final states of $e\mu +$ jets in the $\ttop$ events
where the misidentified $\tauh$ is dominated by the $b$ jet~\cite{CMS:2018jrd}.
In this paper, however,
we take a conservative stance that the $b$ jet has the same misidentification probability
as the other QCD jets~\cite{ATLAS:2017mpa}.
}
\begin{align}
P_{\tau\to \tau}&=0.85, &  
P_{j \to \tau} &=0.02, &   &\hbox{in the one-prong $\tau$ decays;}
\\ \nn
P_{\tau\to \tau}&=0.65, & 
P_{j,b \to \tau}  &=0.01,  & & \hbox{in the three-prong $\tau$ decays.}
\end{align}
We also note that the sign of the electric charge of $\tau_{\rm h}^\pm$
can be determined by the charged tracks.

The $b$--tagging is critical for all three processes. 
We employ  $b$--tagging to remove the $\ttop$ related backgrounds in the
$[\tau\nu][\tau\nu]$ and $[\tau\nu][\tau\nu]WW$
  and to improve the signal preselection for the $[bbW][bbW]$ process.
  In general, $b$--tagging is based on the so-called ghost-association technique \cite{Cacciari:2007fd}
  where a reconstructed jet is $b$--tagged if any $B$ hadron with $p_T> 5~{\rm GeV}$ is found within $\Delta R = 0.3$ of the jet.
In this connection,
we first require that a candidate for a $b$ jet should have minimal acceptance and trigger cuts 
of $p_T> 30\gev$ and $|\eta|<2.5$. 
Then we apply the $b$--tagging efficiency and the mistag rates of the charm or light quark jet as a $b$-jet~\cite{ATL-PHYS-PUB-2016-026,ATL-PHYS-PUB-2017-001}:
\bea
\label{eq:btagging}
P_{b\to b} =70\%,\quad P_{c\to b} = 10\%,\quad P_{j \to b} = 0.2\%.
\eea

For the lepton ($\ell^\pm=e^\pm, \mu^\pm$) identification, we demand the same
rapidity of $|\eta_\ell|<2.5$,
but different $p_T$ cuts for the electron and muon as  $p_T^e > 17\gev$ and $p_T^\mu >15\gev$.
To reduce the leptons from 
decays of heavy hadrons, we apply tight isolation criteria.
For each charged lepton, we compute the isolation variable given by
\bea
I_\ell \equiv \frac{1}{p_T^\ell} \sum_i p_{T_i} \; ,
\eea
where the sum runs over photon, (neutral and charged) hadrons 
within $\Delta R = 0.2~(0.3)$ around the electron~(muon) direction. 
In this analysis, we require $I_\ell<0.06$. 

Finally, we calculate the signal significance including the background uncertainty,
defined by~\cite{Cowan:2010js}
\bea
\label{eq:significance}
\mathcal{S} \!\!\! &=& \!\!\!  
\Bigg[2(N_s + N_b) \log\left(\frac{(N_s + N_b)(N_b + \delta_b^2)}{N_b^2 + (N_s + N_b)\delta_b^2} \right) 
 - 
\frac{2 N_b^2}{\delta_b^2} \log\left(1 + \frac{\delta_b^2 N_s}{N_b (N_b + \delta_b^2)}\right)\Bigg]^{1/2},
\eea
where $N_s$ is the number of signal events, $N_b$ is the number of total background events, 
and $\delta_{b} = \Delta_{\rm bg} N_b$ is the uncertainty on the background yields.

\subsection{$[\tau\nu][\tau\nu]$}
\label{subsec:taunutaunu}

The $[\tau\nu][\tau\nu]$ mode targets at the production of a charged Higgs boson pair,
$pp\to H^+ H^-$,
followed by $\ch\to \tau^\pm_{\rm h}\nu$:
\bea
pp \to H^+ H^- \to \tau^+_h \nu \, \tau^-_h \nu.
\eea
The final state consists of two hadronic $\tau$'s and missing transverse energy.
As shown in Fig.~\ref{fig-production},
this process covers most of the parameter space with $\ma> \mch-10\gev$.

At the 14 TeV LHC
with the total integrated luminosity of $\mathcal{L}_\tot = 3~{\rm ab}^{-1}$,
we prepared the signal samples up to one merged jet
using the MLM scheme based on $k_T$ jet clustering algorithm~\cite{Mangano:2006rw}.
The benchmark point BP--1 in Table~\ref{table:BP:backgrounds} yields
\bea
\label{eq:BP1:signal}
\sigma(pp\to H^+ H^-) + \sigma(pp\to H^+ H^-j) = 0.35 \pb,
\quad \br(\ch\to\tau^\pm\nu)= 0.652,
\eea
where we have imposed $p_T^j > 10\gev$.

Now we cautiously assess the backgrounds.
The backgrounds that we incorporated are based on the samples up to two jets
merged with a parton shower, using the MLM scheme based on $k_T$ jet clustering algorithm~\cite{Mangano:2006rw}.
\bit
\item $pp \to W$+jets where one $\tau_{\rm h}$ comes from $W$ decay and the other $\tau_{\rm h}$
from a jet misidentified as $\tau_{\rm h}$;
\item $pp\to Z/\gm $+jets consisting of $Z/\gamma(\to\tau\tau)$+jets and $Z(\to\nu\nu)$+jets;
\item $\ttop$+jets;
\item $VV'$+jets including $WW$+jets, $WZ$+jets, and $ZZ$+jets.
\item $t W$+jets and $tZ$+jets.
\eit

\begin{table*}[!t]
\setlength\tabcolsep{10pt}
\centering
{\footnotesize\renewcommand{\arraystretch}{1.1} 
\begin{tabular}{c ||c c c c c |c}
\toprule
\multicolumn{7}{c}{$[\tau\nu][\tau\nu]$ }\\
\toprule
 {Cut}   &  $W$+jets & ~~$Z/\gm$+jets~~ & $\ttop$+jets & $VV'$+jets & $N_b$   & $N_s$ \\
 \toprule
Initial & $6.2\times10^{11}$ & $4.39\times10^{10}$ & $1.33\times10^9$ & $4.41\times10^8$ & $6.65\times10^{11}$ & $1.04\times10^6$ \\ 
Basic cuts&  $1.45\times10^7$ & $1.96\times10^8$ & 92929 & 271570 &  $2.11\times10^8$ & 36413\\
$E_T^{\rm miss}>100\gev$ & 298782 & 208799 & 11158 & 14478 &  533217 & 6448\\
$|\Delta\phi(\tau_1,\tau_2)| > 2.4$ & 202117 & 36374 & 5914 & 5503 & 249908 & 3926\\
$\Delta R(\tau_1,\tau_2) < 3$ & 114240 & 8328 & 2926 & 2500& 127994 & 2328\\
$M_{\tau_1\tau_2} > 300 \gev$ & 0 & 1054 & 182 & 183 & 1419 & 465\\
$p_T^{\tau_2} > 100 \gev$ & 0 & 737 & 121 & 121 & 979 & 347\\
$M_T^{\tau_2} > 50 \gev$ & 0 & 0 & 121 & 101 & 222 & 284\\
\bottomrule
\end{tabular}
}
\caption{Cut-flow chart of the number of events of the signal and backgrounds
    for the channel $pp \to H^+ H^- \to [\tau^+_h \nu][\tau^-_h \bar \nu]$
  at the   14 TeV LHC
  with the total integrated luminosity of $\mathcal{L}_\tot = 3~{\rm ab}^{-1}$.
  Negligible backgrounds from $tV$+jets are omitted.
  More details about the selection cuts are in the text.
  }
\label{tab:cutflow:BP1}
\end{table*}

Table \ref{tab:cutflow:BP1} describes the cut-flow of the number of events after
the subsequent selection cuts. 
The ``Basic cuts'' consist of three.
\ben
\item[--]
We veto any event with an electron, a muon, or a $b$-tagged jet. 
Most of the $\ttop$+jets backgrounds are rejected by the $b$-veto.
\item[--] 
We select the events including $\tau_{\rm h}\tau_{\rm h}$,
$\tau_{\rm h}\tau_{\rm h}\tau_{\rm h}$, or $\tau_{\rm h}\tau_{\rm h} j$.
Here a $\tau_{\rm h}$ jet is accepted when $p_T>25\, \gev$ and $|\eta|<2.5$.
\item[--] The electric charges of two $\tau_{\rm h}$ jets should have opposite sign.
\een

\begin{figure}
\includegraphics[width=0.43\textwidth]{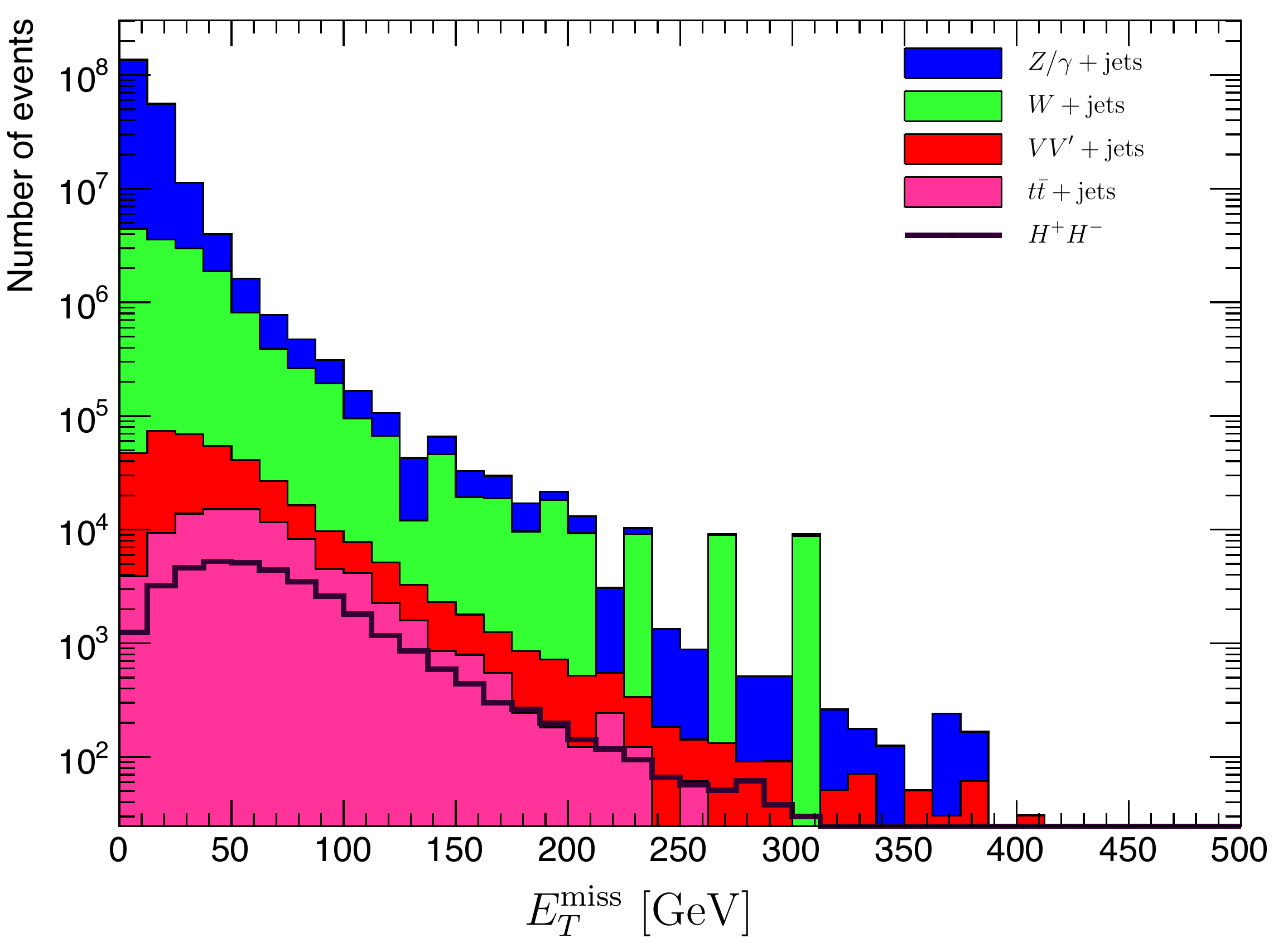}~~
\includegraphics[width=0.43\textwidth]{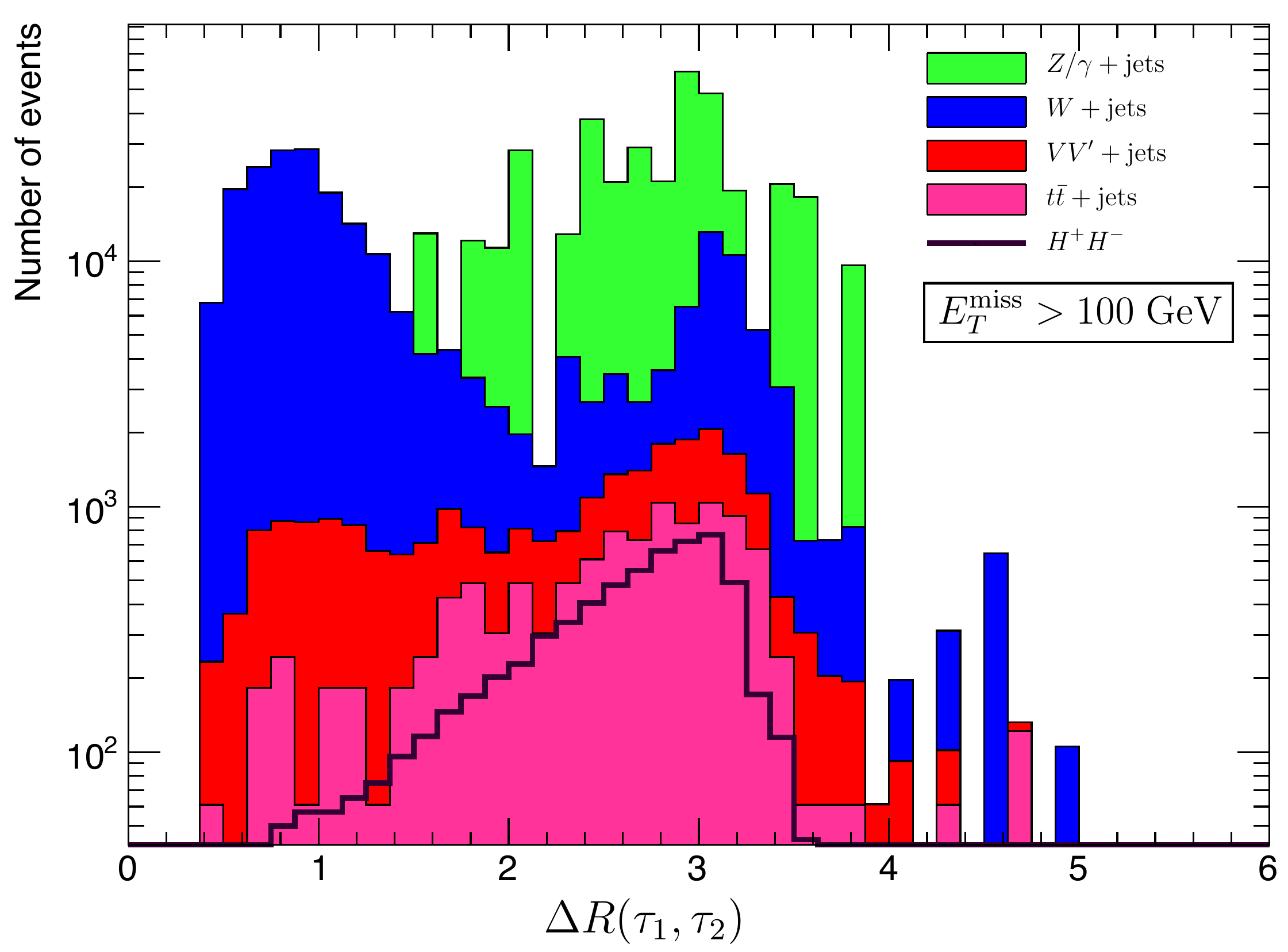}\\
\includegraphics[width=0.43\textwidth]{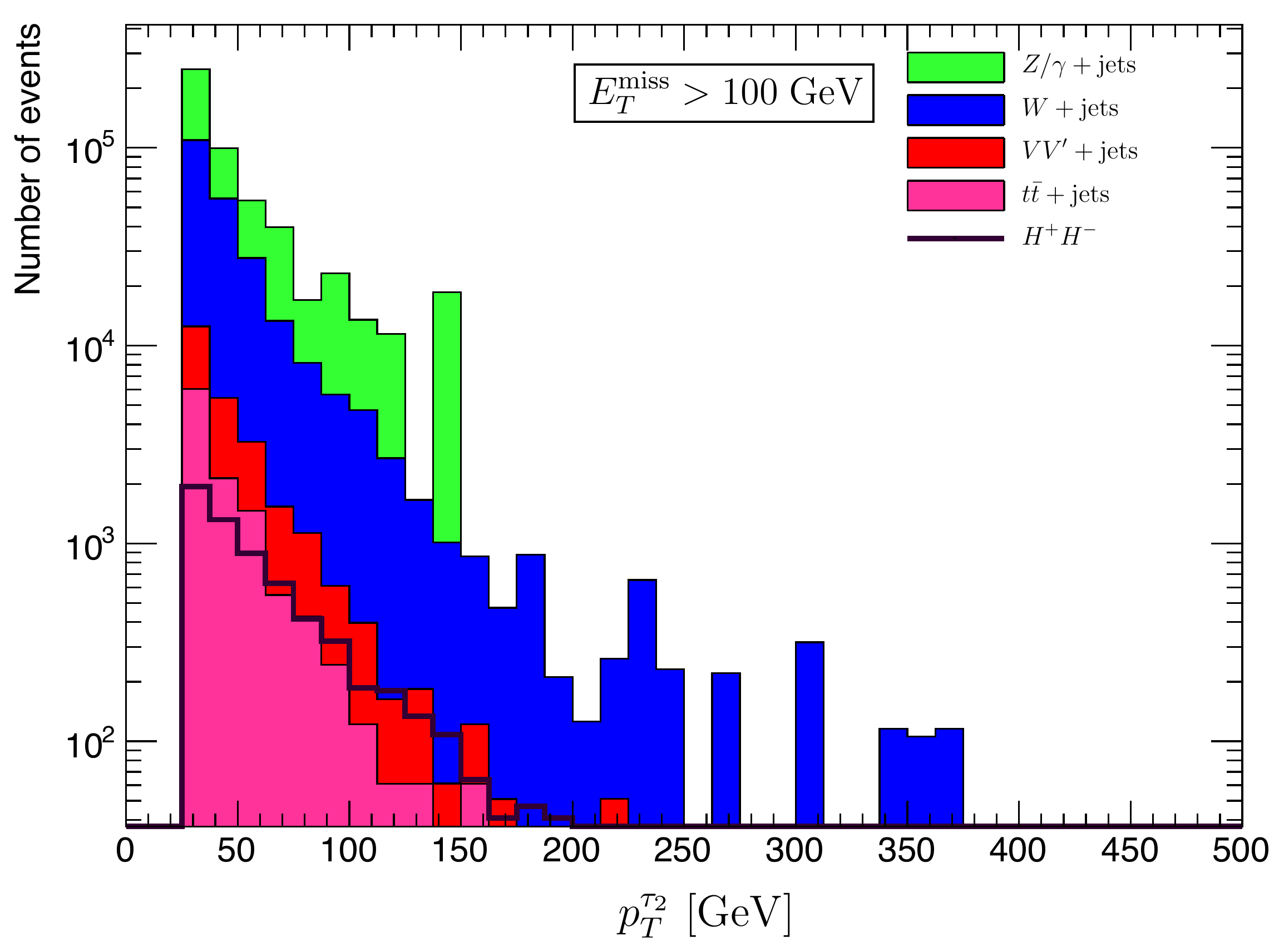}~~
\includegraphics[width=0.43\textwidth]{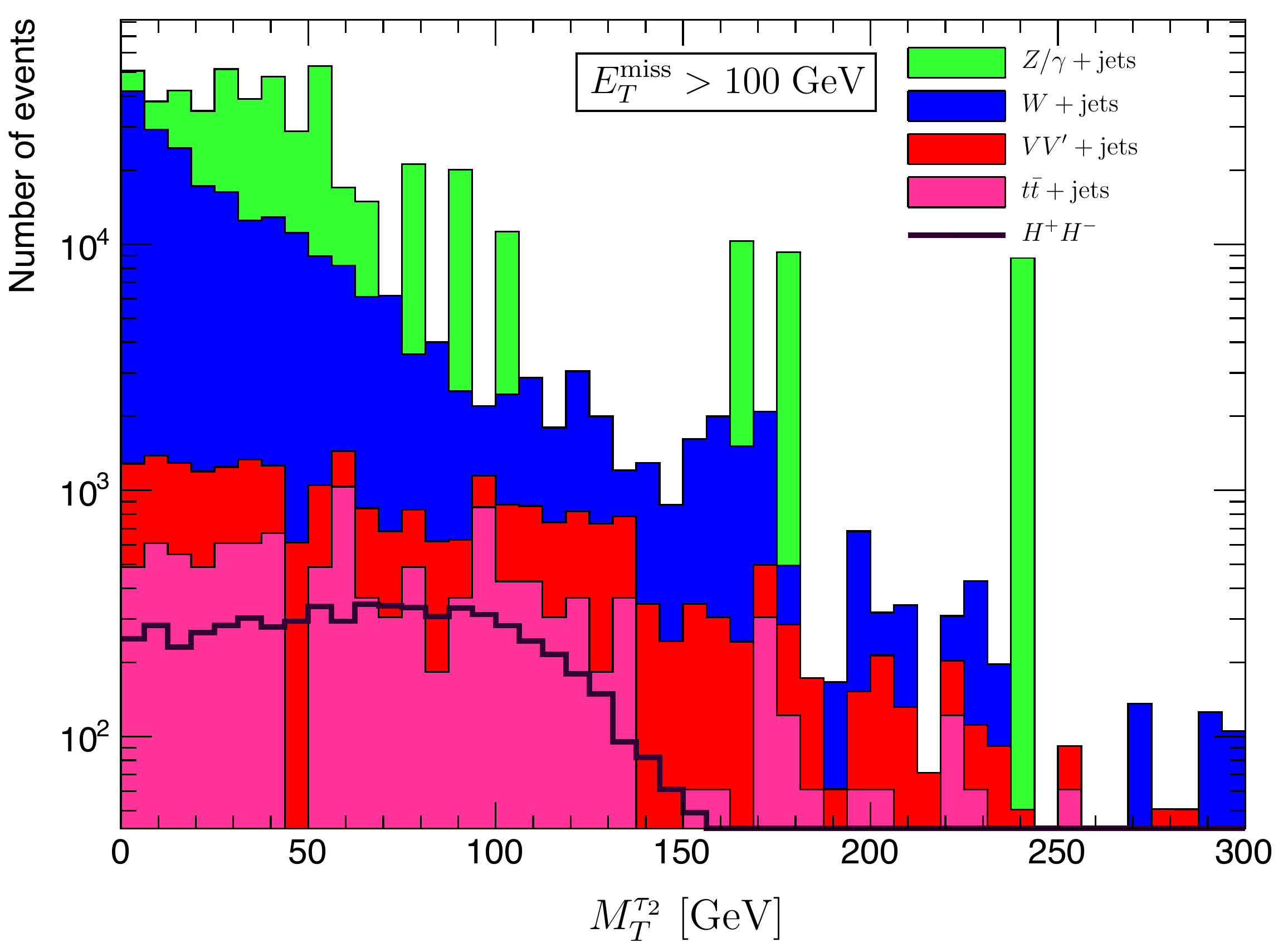}
\caption{
Kinematic distributions for the final state $[\tau\nu][\tau\nu]$ 
about missing transverse energy $E_T^{\rm miss}$ (top-left panel),
$\Delta R(\tau_1,\tau_2)$ (top-right panel),
the transverse momentum of the second-leading tau lepton $p_T^{\tau_2}$ (bottom-left panel), 
and the transverse mass of the second-leading tau lepton $M_T^{\tau_2} $ (bottom-right panel)
at the 14 TeV LHC
with the total integrated luminosity of $\mathcal{L}_\tot = 3~{\rm ab}^{-1}$. 
The different background contributions are stacked on top of
each other, and the expected signal is shown by black line.
The distribution about $\met$ is after imposing the basic cut,
while the others are after imposing $\met > 100\gev$.
}
\label{fig-BP1-distribution}
\end{figure}

For the signal and background events after the basic cuts,
we calculate various kinematic distributions.
We show the distributions about missing transverse energy $E_T^{\rm miss}$ (top-left panel),
$\Delta R(\tau_1,\tau_2)$ (top-right panel),
the transverse momentum of the second-leading tau lepton $p_T^{\tau_2}$ (bottom-left panel), 
and the transverse mass of the second-leading tau lepton $M_T^{\tau_2} $ (bottom-right panel) 
in Fig.~\ref{fig-BP1-distribution}.
$\tau_1$ and $ \tau_2$ are ordered by the $p_T$ such that $p_T^{\tau_1} > p_T^{\tau_2}$.
$M_T^\tau$ is the transverse mass of a tau lepton, defined by
\bea
M_T^\tau = \sqrt{2|p_T^\tau| |E_{T}^{\rm miss}| \times \{ 1 - \cos(\phi_\tau- \phi_{\rm miss})
  \} },
  \eea 
  where $\phi_\tau$ and $\phi_{\rm miss}$ are the
  azimuth angle of the $\tau$ lepton and the missing momentum, respectively.
  As shown in the top-left panel,
$\met$ plays a critical role in separating the signal from the background.
The main backgrounds of $Zjj$ and $Wjj$ yield relatively soft $\met$,
while the signal produces hard $\met$.
For this reason, we enforce $\met>100\gev$.
The other three distributions in Fig.~\ref{fig-BP1-distribution}
are the results after imposing $\met>100\gev$.
Only about 0.2\% of the backgrounds survive the $\met$ selection,
while about 18\% of the signal events remain.

Based on the investigation of the kinematic distributions,
we devise a search strategy summarized in the cutflow.
One of the most efficient cuts is $M_{\tau_1 \tau_2}>300\gev$,
which removes about 99\% of the backgrounds but 80\% of the signal.
In the signal, $M_{\tau_1 \tau_2}$ tends to be high
because two tau leptons originate from different ancestors ($H^+$ and $H^-$).
The cut of $p_T^{\tau_2}>100\gev$ has certain advantage in separating the signal from the backgrounds,
especially $\ttop$+jets and $VV'$+jets:
surviving rate of the signal is about 75\% while that of the total backgrounds is 68\%.
At this level, the dominant background is from $Z/\gm$+jets.
The final selection is on the transverse mass of the second-leading tau lepton,
which aims at a $\tau$ associated with a neutrino.
$M_T^{\tau_2}>50\gev$ removes almost all the backgrounds from $Z/\gm$+jets.
 At the final selection level,
$284$ signal events and $222$ background events survive.
The dominant backgrounds are $\ttop$+jets and $VV'$+jets. 
The significance without the background uncertainty is $19.7$,
which is very promising.
Even with $10\%$ background uncertainty,
the significance is $8.2$.
Certainly, the HL-LHC can probe the light $\ch$ through the $[\tau\nu][\tau\nu]$ final state
if the mass $M_A$ is above the decay threshold of $\ch\to A\wpm$.

\subsection{$[\tau^\pm \nu] [\tau^\pm\nu]W^\mp W^\mp$}

We consider the signal of
\bea
\label{eq:finalstate3}
pp \to H A /HH /AA &&\to  H^- W^+ H^- W^+ + {\rm C.C.}\\ \nn
&&\to \tau_{\rm h}^- \nu \; \ell^+  \nu \;\tau_{\rm h}^- \nu \;\ell^+ \nu+ {\rm C.C.},
\eea
where C.C. denotes the charge conjugate state.
The final state consists of two same-sign leptons, two same-sign hadronic $\tau$'s, and neutrinos. 
We consider the benchmark point BP--2 in Table \ref{table:BP:backgrounds},
where $\br(H \to \ch\wmp) = 0.82$, $\br(A \to\ch\wmp) = 0.88$, and $\br(H^+ \to \tau^+\nu) =0.65$.
Drell-Yan production of $HA$ is dominant over the gluon fusion production of $HH$ and $AA$:
$\sg(\qq \to HA) =75.5\fb$, $\sg(gg \to HH) = 0.68 \fb$, and $\sigma(gg\to AA)=0.80 \fb$.
Therefore, we only generate the signal sample through $\qq \to H A$.

For the final state
$\tau_{\rm h}^-\tau_{\rm h}^-\ell^+ \ell^+ E_T^{\rm miss}$,
the backgrounds are as follows:
\bit
\item $pp \to t\bar{t}+ W^+ \to b \ell^+\nu  \bar{b} \tau^- \nu + \ell^+ \nu $ where one of two $b$ jets or a jet from QCD showering is misidentified as $\tau_{\rm h}$.

\item $pp \to W^-  W^+ W^+  \to  \tau_{\rm h}^- \nu  \ell^+ \nu  \ell^+ \nu $
where a jet from showering is misidentified as $\tau_{\rm h}$.

\item $pp\to Z Z \to \tau^+_{\ell^+}\tau_{\rm h}^-   \tau^+_{\ell^+}\tau_{\rm h}^-$
where $\tau_\ell$ denotes the leptonic decaying $\tau$.

\item $pp \to t\bar{t} + Z \to 
b \ell^+ \nu  \bar{b}\tau^- \nu + \tau^+_{\ell^+} \tau^- $.

\item $pp \to \hsm + Z \to \tau^+_{\ell^+} \tau_{\rm h}^- + \tau^+_{\ell^+}\tau_{\rm h}^-$.
\eit
We also include the backgrounds for the charge conjugate signal state.

\begin{table*}[!t]
\setlength\tabcolsep{10pt}
\centering
{\renewcommand{\arraystretch}{1.1} 
\begin{tabular}{c ||c c c c cc|c}
\toprule
 \multicolumn{8}{c}{$[\tau\nu][\tau\nu]\ell^\pm\nu \ell^\pm\nu$ }\\
\toprule
 {Cut}   &  $\ttop W$ & ~~$WWW$~~ & $ZZ$ & $\ttop Z$ & $\hsm Z$  & $N_b$   & $N_s$ \\
 \toprule 
Initial & 4560 & 1290 & 16567 & 1825 & 1407 & 25649 & 426 \\
Basic cuts & 15.14	& 0.63	& 35.37	& 17.04	& 6.42	& 74.6	& 15.6 \\
$b$-jet veto & 2.7	& 0.62	& 34.97	& 3.42	& 6.35	& 48.06	& 15.43 \\
$E_T^{\rm miss}> 45\gev$ & 2.07	& 0.47	& 7.47	& 2.64	& 2.09	& 14.74	& 10.73 \\
$p_T^{\ell^{\rm (lead)}} < 70\gev $ & 0.94	& 0.19	& 5.33	& 1.53	& 1.43	& 9.42	& 9.59 \\
$p_T^{\tau^{\rm (lead)}} > 40\gev $ & 0.77	& 0.15	& 4.36	& 1.25	& 1.29	& 7.82	& 9.09 \\
$0.4 < \Delta R(\ell, \tau)_1 < 0.8$ & 0.17	& 0.03	& 1.49	& 0.38	& 0.37	& 2.44	& 6.56 \\
$M(\ell, \tau)_1 < 60\gev$ & 0.16	& 0.03	& 1.31	& 0.35	& 0.35	& 2.2	& 6.43 \\
$0.4 < \Delta R(\ell, \tau)_2 < 3.0$ & 0.1	& 0.01	& 1.24	& 0.28	& 0.35	& 1.98	& 6.36 \\
$M(\ell, \tau)_2 < 70\gev$ & 0.04	& 0  	& 1.04	& 0.14	& 0.24	& 1.46	& 6.04 \\ 
\bottomrule
\end{tabular}
}
\caption{Cut-flow chart of the number of events for the final state
  $[\tau^\pm\nu][\tau^\pm\nu]\ell^\mp\nu \ell^\mp\nu$
at the 14 TeV LHC
with the total integrated luminosity of $\mathcal{L}_\tot = 3~{\rm ab}^{-1}$. 
Details about ``Basic cuts'' and the selection are in the text.}
\label{tab:cutflow:taunutaunuWW}
\end{table*}

For event selection, we take the following steps.
The ``Basic cuts" consist of two.
\ben
\item[--] We require two same-sign charged leptons and two same-sign hadronic $\tau$'s
with $p^{\ell,\tau}_T > 20\gev$ and $|\eta_{\ell,\tau}|<2.5$.
\item[--] The electric charge of two same-sign leptons should be opposite to that of two same-sign tau leptons.
\een
After the basic cuts, the signal rate is considerably reduced.
The resulting acceptance times efficiency, $\mathcal{A}\times \es$, is about $3\%$. 
But the reduction of the total backgrounds is more severe
with $\mathcal{A}\times \es \simeq 0.3\%$.
The basic cuts are most effective in the $WWW$ background process
since it is difficult for a QCD showering jet (mistagged as $\tau_{\rm h}$) to satisfy the requirement for the $p_T$ and electric charge.
The second selection is the $b$-jet veto.
We reject the event including any $b$-tagged jet with $p^b_T > 30\gev$ and $|\eta_b|<2.5$.
It is designed to suppress the $t\overline{t}W$ and $t\overline{t}Z$ backgrounds,
which results in a roughly 80\% cut.
On the contrary, the events from signal and other backgrounds remain almost intact. 
At this level, the significance without the background uncertainty is about 2.

\begin{figure}
\centering
\includegraphics[width=0.44\textwidth]{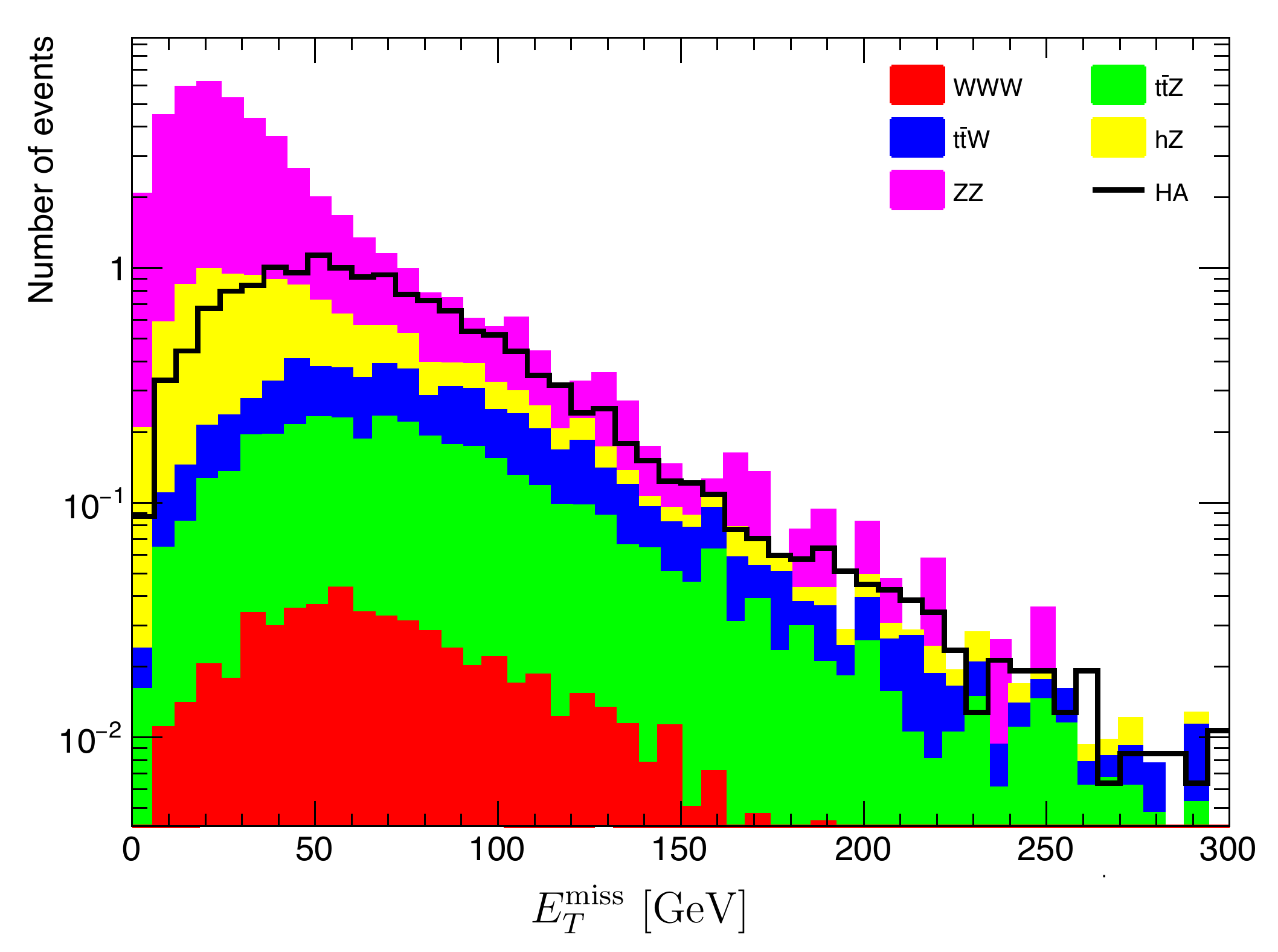}
\includegraphics[width=0.44\textwidth]{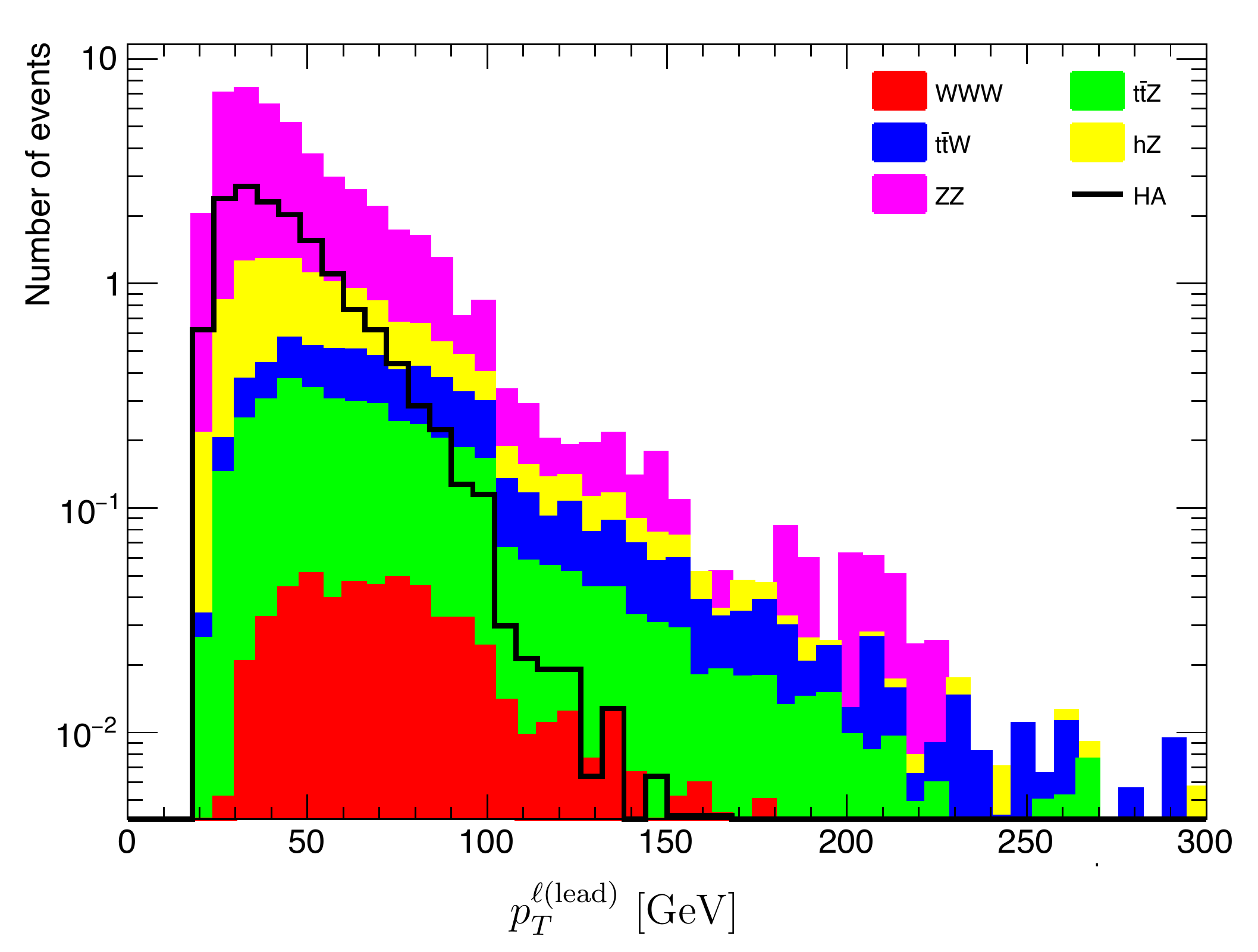}
\includegraphics[width=0.44\textwidth]{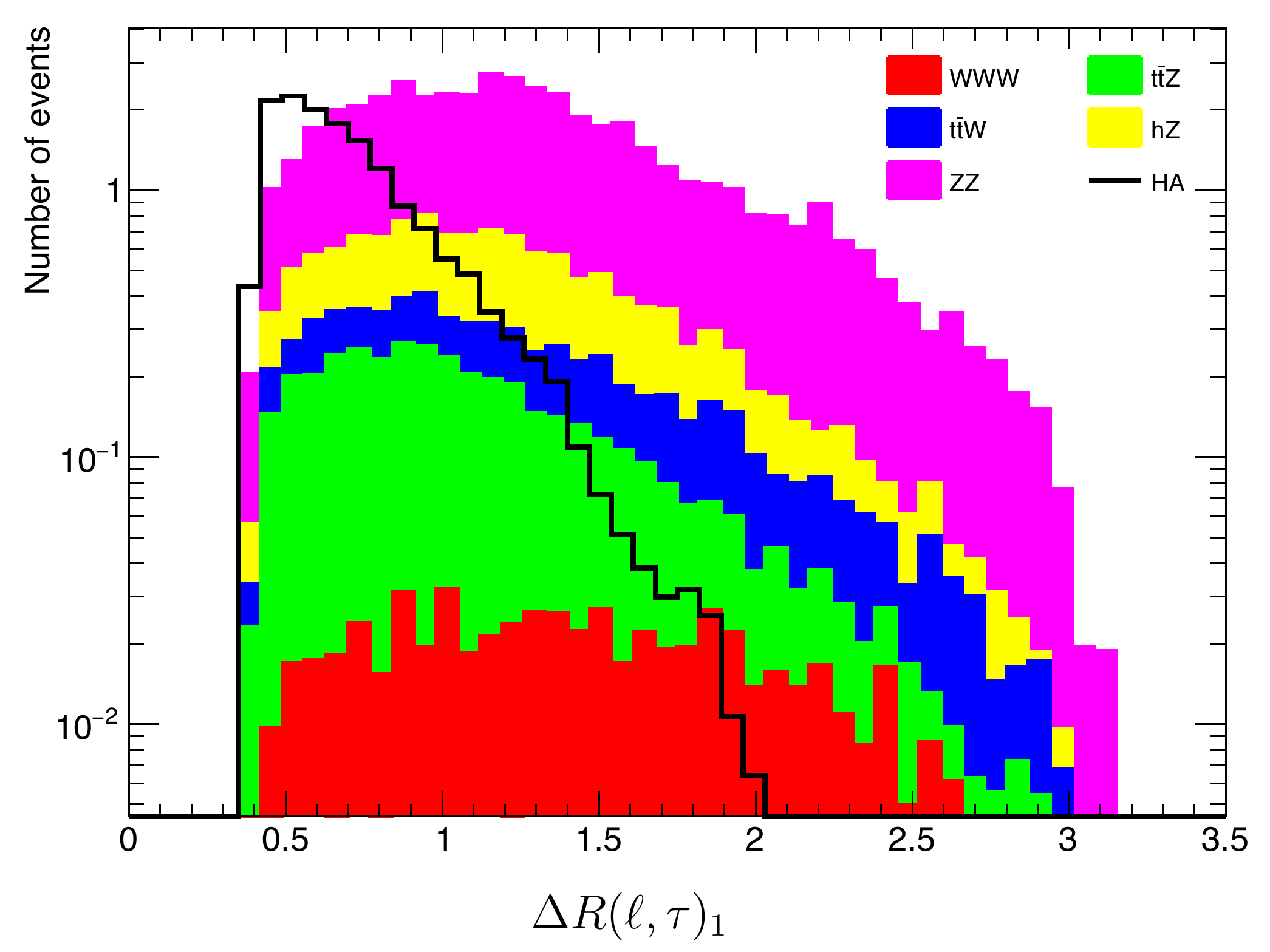}
\includegraphics[width=0.44\textwidth]{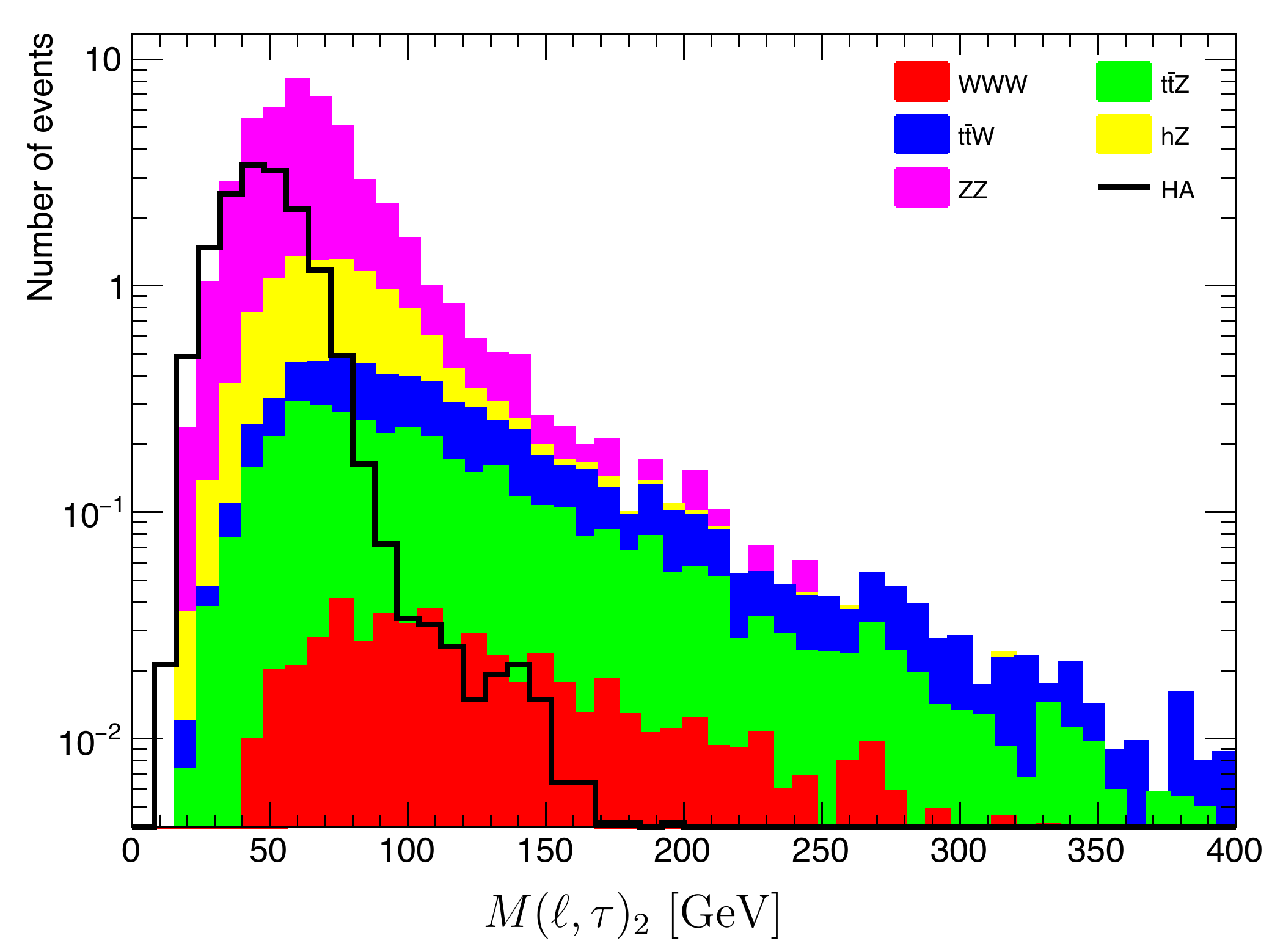}
\caption{
Kinematic distributions for the final state $[\tau\nu][\tau\nu]\ell^\pm\nu \ell^\pm\nu$ at the 14 TeV LHC
with the total integrated luminosity of $\mathcal{L}_\tot = 3~{\rm ab}^{-1}$. 
The results shown are after the basic cuts and $b-$jet veto.
The different background contributions are stacked on top of
each other, and the expected signal is shown by black line.
Four representative distributions are presented
on the missing transverse energy $E_T^{\rm miss}$ (top-left panel),
the leading charged lepton transverse momentum $ p_T^{\ell \rm (lead)}$ (top-right panel),
the angular separation between the lepton and the $\tau$ in the $(\ell,\tau)_1$ pair (bottom-left panel),
and the invariant mass of the lepton and the $\tau$ in the $(\ell,\tau)_2$ pair
(bottom-right panel).
}
\label{fig-taunutaunuWW}
\end{figure}

To devise more sophisticated selections,
we show in Fig.~\ref{fig-taunutaunuWW} the kinematic distributions of the signal and backgrounds
about missing transverse energy $E_T^{\rm miss}$ (top-left panel),
the transverse momentum of the leading lepton $p_T^{\ell^{(\rm lead)}}$ (top-right panel),
the angular separation $ \Delta R(\ell, \tau)_1 $ (bottom-left panel), 
and the invariant mass $ M(\ell, \tau)_2 $ (bottom-right panel).
The results are based on the events passing the basic cuts and $b-$jet veto.
The first decisive cut is from the $E_T^{\rm miss}$ distribution.
Both $ZZ$ and $\hsm Z$ backgrounds have lower $E_T^{\rm miss}$ than the signal.
We take $E_T^{\rm miss}>45\gev$ as the third selection, removing about 70\% of $ZZ$ and $\hsm Z$ backgrounds.
The next important selection comes from the $p_T^{\ell^{(\rm lead)}}$ distribution (top-right panel)
 where $\ell^{(\rm lead)}$ denotes the lepton with the largest $p_T$.
$p_T^{\ell^{(\rm lead)}}$ in the signal is softer than that in most of the backgrounds,
while $p^{\tau ^{\rm (lead)}}_T$ in the signal is relative harder. 
In this regard, we select the events with $p_T^{\ell^{(\rm lead)}} < 70\gev$ 
and $p_T^{\tau^{(\rm lead)}} > 40\gev$.

The final four selections in Table \ref{tab:cutflow:taunutaunuWW}
are motivated by the characteristic of the signal
$pp \to H + A \to H^+_{ \tau^+\nu}W^-_{\ell^-\nu} +  H^+_{ \tau^+\nu}W^-_{\ell^-\nu} $.
Two same-sign charged leptons in the signal come from different mother particles, $H$ and $A$.
To make the best use of the feature,
we first select a pair of $\ell$ and $\tau$ with minimal $\Delta R(\ell_i, \tau_j)$, and call the pair $(\ell,\tau)_1$.
The remaining pair of lepton and $\tau$ is $(\ell,\tau)_2$.
In the bottom-left panel of Fig.~\ref{fig-taunutaunuWW},
we show the distribution of the angular distance between the
lepton and $\tau$ inside $(\ell, \tau)_1$.
The bottom-right panel presents the distribution of the invariant mass of
the lepton and $\tau$ inside $(\ell, \tau)_2$.
The signal is mainly populated in the regions of low $ \Delta R(\ell, \tau)_{1,2} $ and low $M(\ell, \tau)_{1,2}$,
compared with the backgrounds.
So we make the final four selections of 
$0.4 < \Delta R(\ell, \tau)_1 < 0.8$, $ M(\ell, \tau)_1 < 60\gev $, 
$0.4 < \Delta R(\ell, \tau)_2 < 3.0$, and $ M(\ell, \tau)_2 < 70\gev $.  
They are efficient to control the whole backgrounds,
especially the $t\overline{t}W$, $t\overline{t}Z$, and $WWW$ backgrounds.
As $2\rightarrow 3$ scattering processes,
these backgrounds yield wide opening angles, which fail the final four selections.
The $ZZ$ and $\hsm Z$ backgrounds 
also prefer wide opening angles because of lighter masses of $Z$ and $\hsm$ than $H$ and $A$.
About $80\%$ of these two backgrounds are removed.

After all of the above selections, $6.04$ signal events and $1.46$ background events (mostly $ZZ$ background) 
are left. 
The significance without the background uncertainty is $3.53$. 
If we include $10\%$ background uncertainty,
the significance slightly reduces to $3.48$.
Marginally, the HL-LHC can probe the light $\ch$ through
  the signal of two same-sign charged leptons and two same-sign
  hadronic $\tau$'s  in the $[\tau\nu][\tau\nu]WW$ final state.

\subsection{$[bbW][bbW]$}

The $[bbW][bbW]$ process targets the production of a pair of charged Higgs bosons, followed by $\ch\to A W^{\pm(*)}$
\bea 
p p (q\bar{q}/gg) \to H^+ H^- \to 
A W^+ A W^-
\to  b\bar{b} \ell^+ \nu_\ell b\bar{b} q\bar{q}' + {\rm C.C.}
\eea
We consider the benchmark point BP--3 in Table \ref{table:BP:backgrounds} where the cross sections for the signal process are
\bea
\hbox{BP--3: } && \sg(\qq \to H^+ H^-) = 185.4\fb, \quad \sg(gg\to H^+ H^-) = 25.6\fb
.
\eea
The cross section of loop-induced gluon fusion production 
is about 10\% of the
Drell-Yan production cross section.
For the decay of $WW$, we consider the semi-leptonic decays.
Then the backgrounds are as follows: 
\bit
\item $pp \to \hsm(\to \bb) V + {\rm jets}$ where the jets from the QCD showering are misidentified as $b$ jets;
\item $pp \to tV + \ttop \hsm + \ttop V$;
\item $VV'+ {\rm jets}$;
\item $ZZ+\bb$;
\item $\ttop+ {\rm jets}$.
\eit

\begin{table*}[!t]
\setlength\tabcolsep{6pt}
\centering
{\footnotesize\renewcommand{\arraystretch}{1.1} 
\begin{tabular}{l|| cccc c |c}
\toprule
 \multicolumn{7}{c}{$[\bb W][\bb W]$ }\\
\toprule
 {Cut}   & $\hsm V$+jets & $tV+t\bar{t}\hsm/V$ & $VV+ZZbb$ & $t\bar{t}+{\rm jets}$ & $N_b$ & $N_s$ \\
 \toprule
Initial & $6.09 \times 10^6$ & $97.4 \times 10^6$ &   $440.9 \times 10^6$ & $1.34 \times 10^9$ & $1.90\times 10^9$  &  $6.33 \times 10^5$  \\
Basic cuts  & $4.27 \times 10^5$ &   $15.1 \times 10^6$ &   $46.42 \times 10^6$ &  $206.42 \times 10^6$  &  $2.69 \times 10^8$ & $8.01 \times 10^4  $ \\
$N_{\rm jets} \geq 4, N_{b} \geq 2$ & $1.06 \times 10^4$ &   $1.0 \times 10^6$ &   $1.52 \times 10^5$ &   $55.56 \times 10^6$ &  $ 5.67 \times 10^7$ &   $1.06 \times 10^4$  \\
$M_T^W < 150~{\rm GeV}$ & $1.04 \times 10^4$ &   $9.65 \times 10^5$ &  $1.45 \times 10^5$ &   $54.11 \times 10^6$ &  $5.54 \times 10^7$ &   $1.03 \times 10^4$\\
$M_{b_1 b_2} < 100~{\rm GeV}$ & $5.86 \times 10^3$ &  $4.59 \times 10^5$ &   $1.13 \times 10^5$ &   $20.29 \times 10^6$ &  $2.09\times 10^7$  &    $7.21 \times 10^3$ \\
$\ptll < 350~{\rm GeV}$ & $5.85 \times 10^3$ &   $4.59 \times 10^5$  &  $1.13 \times 10^5$ &   $20.28 \times 10^6$ &  $2.08\times 10^7$ &    $7.21 \times 10^3$ \\
$p_T^{\rm jet} < p_T^{j,\rm max}$ & $5.81 \times 10^3$ &   $4.56 \times 10^5$ &   $1.10 \times 10^5$ &   $20.18 \times 10^6$  & $2.08 \times 10^7$ &  $7.11 \times 10^3$    \\
$E_{T}^{\rm miss} < 0.7 H_T$ & $5.72 \times 10^3$ &   $4.49 \times 10^5$ &   $1.09 \times 10^5$ &   $20.00 \times 10^6$ &   $2.06 \times 10^7$ &   $7.06 \times 10^3$ \\
top  veto & $5.11 \times 10^3$  &   $4.14 \times 10^5$ &   $1.00 \times 10^5$ &   $18.67 \times 10^6$ &   $1.92 \times 10^7$  &  $6.78 \times 10^3$ \\
cuts on $M_{bb jj}$ and $ M_{bb} $ 
& $2.20 \times 10^2$  &   $1.45 \times 10^4$ &    $2.49 \times 10^3$ &   $6.08 \times 10^5$ &  $6.25 \times 10^5$   &   $3.90 \times 10^2$ \\
$H_T < 400~{\rm GeV}$ & $1.92 \times 10^2$ &   $1.21 \times 10^4$ &   $1.84 \times 10^3$ &   $5.18 \times 10^5$ &  $5.33 \times 10^5$   &  $3.08 \times 10^2$ \\
\midrule
$N_b = 3$ & $3.2 \times 10^1$  &    $1.16 \times 10^3$ &   $1.54 \times 10^2$ &   $7.12 \times 10^4$ & $7.25 \times 10^4$  &  $5.73 \times 10^1$ \\
$N_b = 4$ & $0$ &  $1.40 \times 10^2$ & $0$ &   $6.08 \times 10^3$ &  $6.23 \times 10^3$  &  $1.42 \times 10^1$ \\
\bottomrule\end{tabular}
}
\caption{Cut-flow chart of the number of events of the signal and backgrounds
for the channel  $[\bb W][\bb W]$   at the 14 TeV LHC
with the total integrated luminosity of $\mathcal{L}_\tot = 3~{\rm ab}^{-1}$.
Details about the selections are in the text.}
\label{tab:cutflow:bbWbbW}
\end{table*}

\begin{figure}[!t]
\centering
\includegraphics[width=0.43\textwidth]{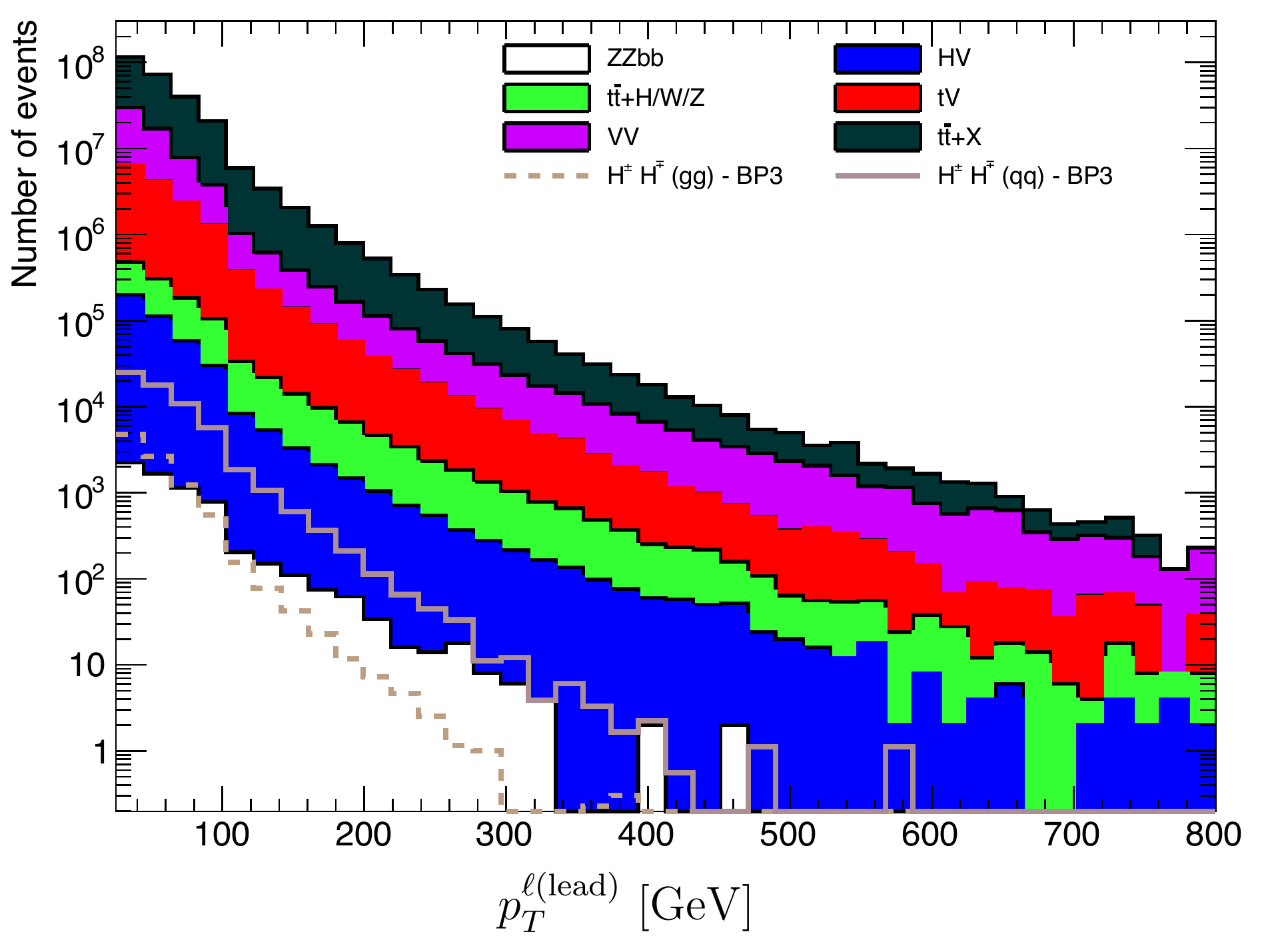}~~
\includegraphics[width=0.43\textwidth]{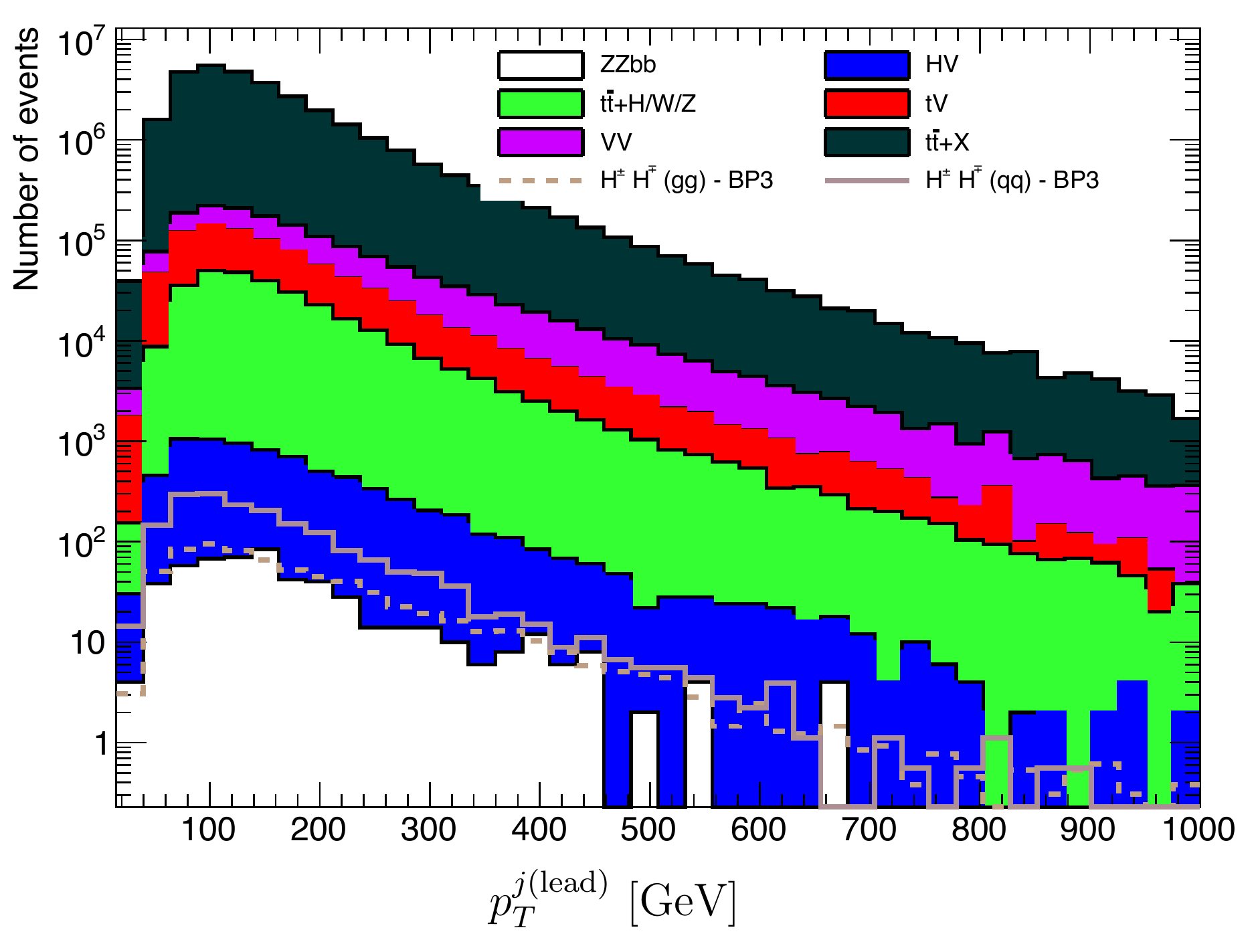}
\\
\includegraphics[width=0.43\textwidth]{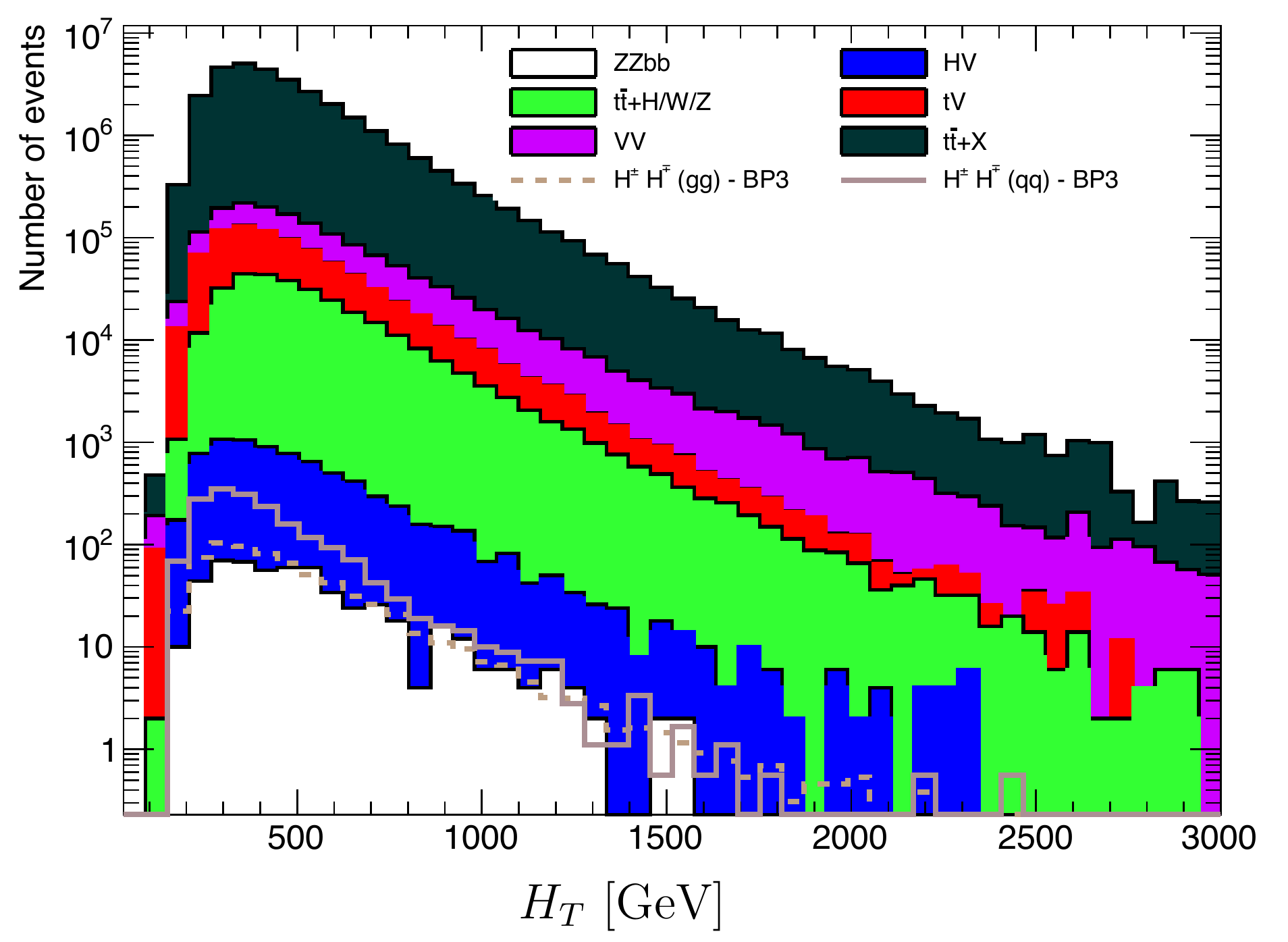}~~
\includegraphics[width=0.43\textwidth]{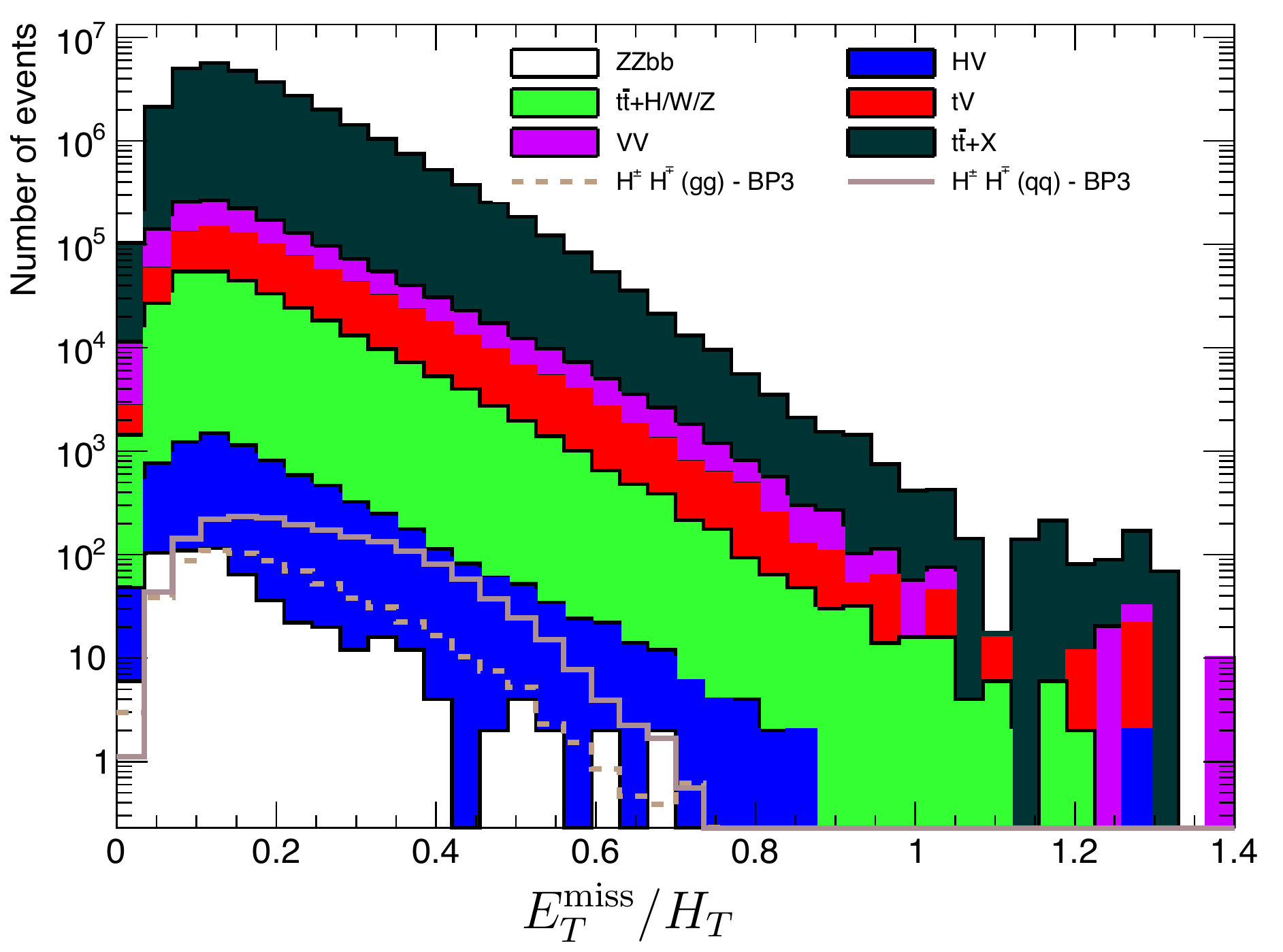}
\vspace{-0.4cm}
\caption{
  Examples for few selected distributions which we used in the
  signal-to-background optimization analysis: the
  leading jet transverse momentum (top-left panel); 
  the leading lepton transverse momentum (top-right panel);
 $H_T$, the scalar sum of
  jet transverse momenta (bottom-left panel); 
  the ratio of missing transverse
  energy to $H_T$ (bottom-right panel). The backgrounds shown here correspond to
  $ZZbb$ (white), $HV$ (blue), $t\bar{t}+H/W/Z$ (green), $tV$ (red),
  $VV$ (magenta) and $t\bar{t}+{\rm jets}$ (dark green). In the same
  canvas, we show the $gg\to H^+ H^-$ (dashed line) and $q\bar{q} \to
  H^+ H^-$ (solid line) for BP--3 (light sienna).
  }
\label{fig:dist:bbWbbW}
\end{figure}

At the LHC, $[bbW][bbW]$ is the most challenging process to probe the light $\ch$ in type-I.
The signal significance at the final selection shall be shown to be very small, far below the discovery level.
Nevertheless, 
we present our investigation of all the available kinematic distributions and the effects of various kinematic cuts,
hoping they help the future study. 
Some of the key
distributions are shown in Fig.~\ref{fig:dist:bbWbbW}.  
First, we apply the \enquote{Basic cuts}, consisting of three selections.
\ben
\item[--] We select events if they contain exactly one charged lepton with
$p_T^\ell > 25~{\rm GeV}$ and $|\eta| < 2.5$. 
\item[--] We apply a veto on hadronically decaying tau leptons. Events
should not contain any $\tau_h$ with $p_T > 20\gev$.
\item[--] $\met > 30\gev$.
\een
After the basic cut, the signal significance without the background uncertainty is 4.81.
As soon as we include the background uncertainty, 
however, the significance drops quickly, e.g., into 0.03 with $\Dt_{\rm bg}=1\%$.
We need to reduce the background events.

We impose the cuts on the number of jets and $b$ jets, 
$N_j \geq 4$ and $N_b\geq 2$,
which is a key discriminator between the signal and background. 
In the signal, the number of jets is at least six 
wherein four of them are $b$ jets.
But a large portion of the $b$ jets from a light $A$ are too soft to pass the jet
selection threshold.
Therefore, we impose a looser jet selection such that the events contain at least four jets and at least two $b$--jets. 
This selection, by itself, reduces the number of $t\bar{t}$ events by a factor of 4 and
the signal by a factor of 8.

The charged lepton in the signal comes from the $W$ decay.
To take the full advantage of the feature,
we pair the charged lepton with the missing transverse energy
and construct the transverse mass $M_T^W$, defined by
\bea 
M_T^W = \sqrt{2|p_T^\ell|
  |E_{T}^{\rm miss}| \times (1 - \cos\Delta\phi)},
\eea 
where $\Delta\phi = \phi_\ell - \phi_{\rm miss}$. 
We require $M_T^W < 150~{\rm GeV}$.
But it is not efficient since the charged lepton in the backgrounds involving top quarks
comes from $W$ also.
Another characteristic of the signal is that two $b$ jets come from a common ancestor.
In the background, they are from different ancestors.
We impose a condition on the invariant mass of the
leading $b$ jet ($b_1$) and the subleading $b$ jet ($b_2$)
such that $M_{b_1 b_2} < 100~{\rm GeV}$.
This selection is effective to suppress the $t\bar{t}+{\rm jets}$ (from $\sim 54$M events to $\sim 20$M events).

Further requirements are on the transverse momenta of the charged lepton and
jets. First, we select events if the transverse momentum of the leading charged
lepton is smaller than $350~{\rm GeV}$. Second, we
require that the transverse momentum of a jet be
smaller than $p_T^{j,\rm max}$,
defined by $p_T^{j,\rm max} = 500, 350, 250,
150~{\rm GeV}$ for the leading, subleading, third, and fourth jet, respectively, 
regardless of whether they are $b$--tagged or not. 
Unfortunately, the cuts on $p_T^{\ell,j}$ hardly separate the signal from the backgrounds
since basically the shapes of the $p_T^{\ell,j}$ distributions are very similar: see the top panels in Fig.~\ref{fig:dist:bbWbbW}.

Now we investigate the scalar sum of transverse momenta of jets,
defined by 
\bea 
H_T = \sum_{i \in {\rm jets}} p_T^i,
\eea
of which the distributions for the signal and backgrounds are in the bottom-left panel of Fig.~\ref{fig:dist:bbWbbW}.
The background processes produce a hard $H_T$ spectrum, 
while the signal has a softer spectrum.
With the hope that some correlations of various energy observations (such as $\met$ and the effective mass $M_{\rm eff}$)
to $H_T$ may suppress the backgrounds, 
we examine the distributions of ${E_{T}^{\rm miss}}/{H_T}$, $(E_{T}^{\rm miss} + p_T^\ell)/{H_T}$, 
${E_{T}^{\rm miss}}/{M_{\rm eff}}$, and $ \lf {E_{T}^{\rm miss}+p_T^\ell}\ri/{M_{\rm eff}}$.
Here $M_{\rm eff} = p_T^\ell + E_T^{\rm miss} + H_T$.
Since all of them give almost the same results,
we choose $\met/H_T<0.7$ as a representative: see the bottom-right panel of Fig.~\ref{fig:dist:bbWbbW}.

Since the backgrounds involving $\ttop$ are still dominant,
we further apply a top quark veto.
Aiming at hadronically decaying top quark candidates, we construct $W$-candidates from any two
jets with $p_T > 25~{\rm GeV}$ and $\Delta R(j_1, j_2) < 1.5$. Then,
we veto events if any additional $b$ jet with $p_T > 30~{\rm GeV}$ and
$\Delta R(W_{jj}, b) < 1.5$ satisfies
\bea
X_{tt} \equiv \sqrt{\bigg(\frac{M_{jj} - M_W}{0.1 M_{jj}} \bigg)^2 + \bigg(\frac{M_{jjb} - m_t}{0.1 M_{jjb}} \bigg)} < 3.2,
\eea
for any possible combination. 
Next, we select hadronically decaying charged Higgs candidates. 
First, we construct two dijet systems,
$jj$ from the decay of the $W$-boson and $bb$ from the decay of $A$. 
The dijet is formed if two jets are within $\Delta R < 1.5$. 
The two dijets, $jj$ and $\bb$, are then combined to form a charged Higgs candidate,
while the $\bb$ dijet system is to form $A$. 
Combining these, we require
\bea 
|M_{jjbb} - M_{H^\pm}| < 10~{\rm GeV}, \qquad |M_{bb} - M_A| < 10~{\rm GeV}.
\eea 
Finally we demand that the $H_T$ variable be smaller than $400~{\rm GeV}$. 

The last discussion is on categorizing the events according to the number of $b$ jets. 
At the end of the selection ($H_T<400\gev$), 
236 signal events remained in the $N_b = 2$ region, 
$57$ events in the $N_b = 3$ region and $14$ events in the $N_b = 4$ region. 
The resulting significance is very small, about $0.2$--$0.4$ after these selections. 
It turns out that more work is needed to refine the selection and to enhance the significance 
using deep-learning algorithms for example. 
Furthermore, we expect better perspective at electron-positron colliders at $\sqrt{s} = 250~{\rm GeV}$ 
where we expect almost background free environment, thanks to the absence of $t\bar{t}$ backgrounds.

\section{Conclusions}
\label{sec:conclusions}

In the framework of the type-I 2HDM,
we have comprehensively studied the phenomenology to set a full roadmap for the light charged Higgs boson. 
The existing constraint of $b \to s \gamma$ puts a very stringent lower bound
  on the mass of the charged Higgs boson in type-II and type-Y, 
as $\mch\gsim 800\gev$~\cite{Misiak:2020vlo}.  
That is why we focus on type-I here.
Imposing the light mass for the charged Higgs boson severely limits type-I,
even without any assumptions on the model parameters, 
because of existing electroweak precision data, Higgs data, $b \to s \gamma$,
and direct searches at the LEP, Tevatron, and LHC.
The masses and couplings of the other Higgs bosons, $A$ and $H$, 
are considerably restricted:
(i) $\ma$ and $\mhh$ are below about 570 GeV;
(ii) there is a significant correlation between $\ma$ and $\mhh$
(e.g., $\ma$ and $\mhh$ cannot be simultaneously heavy);
(iii) a light $\ma$ allows small $\tan\beta$;
(iv) the current data still permit substantial deviation from the Higgs alignment.

We rummaged among the finally allowed parameter space 
and found that the critical parameter is the mass of the pseudoscalar Higgs boson $A$.
When $A W^\pm$  is beyond the decay threshold of the charged Higgs boson, 
$H^\pm$ decays into
a fermion pair, mainly into $\tau^\pm \nu$.
Since only large $\tan\beta(\gsim 10)$ is allowed when $\ma$ is heavy,
the conventional production channel in the search for the light $\ch$,
via the top quark decay, is not helpful.
We found that the pair production of charged Higgs bosons
has higher discovery potential.
The associated production of $H$ and $A$,
followed by $H/A \to \ch\wmp$,
is also efficient to probe the light $\ch$.
When $A W^\pm$  is below the decay threshold of the charged Higgs boson, 
$H^\pm$ will mostly decay into $A W^\pm$
and $A$ into $\bb$.
Based on these characteristics, we assessed  the detection significance of light charged Higgs bosons 
in three final states,
$[\tau \nu] [ \tau \nu]$,
$[\tau \nu] [ \tau \nu] \ell^\pm \nu \ell^\pm \nu$, and
$[b\bar b W ] [b \bar b W]$.
While we enjoy a large significance for the first final state and a reasonable
significance for the second final state, the last one suffers from huge $t\bar t$ related
backgrounds. 

Before we close, a few comments are offered as follows:
\begin{enumerate}
\item The decay of the charged Higgs boson into $A W^\pm$ depends on the gauge coupling,
  which is independent of Yukawa couplings, in contrast to fermionic decays.  
  Once kinematically allowed, therefore, $\ch\to A W^\pm$ dominates over the fermionic
  modes.  Thus, the mass of the pseudoscalar Higgs boson is a crucial factor in searching
  for the charged Higgs boson. 

\item The golden channel for the light Higgs boson
is  $pp \to H^+ H^- \to [\tau^+\nu ]\,[ \tau^- \bar \nu]$ when the decay into $A\wpm$ is
  kinematically suppressed.  
  Its signal rate enjoys a large significance. 
  The benchmark point BP--1 that we illustrated gives a typical size of the cross sections in the
  allowed parameter space, thus the significance of other allowed parameter sets would
  not be substantially different. 
  New techniques for improving the $\tau$--tagging using multivariate discriminants and the measurement
of the $\tau$ charge will further enhance the significance.
  
\item On the other hand, when the decay into $A \wpm$ is kinematically allowed,
  the decay chain $pp \to H^+ H^- \to [ A W^+] \, [A W^- ] \to [b\bar b \ell^+\nu ] \,
  [b\bar b qq]$ is dominant in type-I but suffers the  huge background from $t\bar t +$jets.  For this case, we do not 
  get any significant sensitivity. We notice that this channel can be tested in the future electron-positron colliders at center-of-mass energy of $250$ GeV due to the absence of the $t\bar{t}$ backgrounds.
  
  \end{enumerate}

\acknowledgments
K.C. would like to thank Professor Wai-Yee Keung of University of Illinois at Chicago 
for hospitality.
K.C. was supported by MoST with grant numbers MoST-107-2112-M-007-029-MY3 and
MOST-110-2112-M-007-017-MY3. 
The work 
of AJ is supported by a KIAS Individual Grant No. QP084401 via the Quantum 
Universe Center at Korea Institute for Advanced Study.
The work of JK, SL, and JHS is supported by 
the National Research Foundation of Korea, Grant No.~NRF-2019R1A2C1009419. C.T.L. is supported by KIAS Individual Grant No. PG075301 (C.T.L.).

\end{document}